\documentclass[pre,twocolumn,amsmath,amssymb,nofootinbib,floatfix,superscriptaddress]{revtex4}

\usepackage{graphicx,bm,xcolor, bbold}
\usepackage{enumerate}
\usepackage{graphicx,bm}
\usepackage[normalem]{ulem}

\makeatletter
\def\graphicscale{\twocolumn@sw{0.3}{0.4}}
\def\graphicthreescale{\twocolumn@sw{0.3}{0.4}}

\begin{document}

\title{Uncovering gauge-dependent critical order-parameter
  correlations \\ by a stochastic gauge fixing at O($N$)$^*$ and
  Ising$^*$ continuous transitions}

\author{Claudio Bonati} \affiliation{Dipartimento di Fisica
  dell'Universit\`a di Pisa and INFN Sezione di Pisa, Largo Pontecorvo
  3, I-56127 Pisa, Italy}

\author{Andrea Pelissetto}
\affiliation{Dipartimento di Fisica dell'Universit\`a di Roma Sapienza
        and INFN Sezione di Roma I, I-00185 Roma, Italy}

\author{Ettore Vicari} 
\affiliation{Dipartimento di Fisica dell'Universit\`a di Pisa,
        Largo Pontecorvo 3, I-56127 Pisa, Italy}

\date{\today}

\begin{abstract}
We study the O($N$)$^*$ transitions that occur in the 3D ${\mathbb
  Z}_2$-gauge $N$-vector model, and the analogous Ising$^*$
transitions occurring in the 3D ${\mathbb Z}_2$-gauge Higgs model,
corresponding to the ${\mathbb Z}_2$-gauge $N$-vector model with
$N=1$.  At these transitions, gauge-invariant correlations behave as
in the usual $N$-vector/Ising model.  Instead, the nongauge invariant
spin correlations are trivial and therefore the spin order parameter
that characterizes the spontaneous breaking of the O($N$) symmetry in
standard $N$-vector/Ising systems is apparently absent. We define a
novel gauge fixing procedure---we name it stochastic gauge
fixing---that allows us to define a gauge-dependent vector field that
orders at the transition and is therefore the appropriate order
parameter for the O($N$) symmetry breaking.  To substantiate this
approach, we perform numerical simulations for $N=3$ and $N=1$. A
finite-size scaling analysis of the numerical data allows us to
confirm the general scenario: the gauge-fixed spin correlation
functions behave as the corresponding functions computed in the usual
$N$-vector/Ising model. The emergence of a critical vector order
parameter in the gauge model shows the complete equivalence of the
O($N$)$^*$/Ising$^*$ and O($N$)/Ising universality classes.
\end{abstract}

\maketitle

\section{Introduction}
\label{intro}

Gauge symmetries and Higgs phenomena are key features of theories
  describing collective phenomena in condensed-matter
  physics~\cite{Anderson-book,Wen-book,Fradkin-book,Sachdev-19}.  To
  understand these phenomena, and, in particular, the major mechanisms
  driving phase transitions and critical phenomena in these theories, it is
  crucial to achieve a solid understanding of the interplay between
  global and gauge symmetries, and, in particular, of the role that
  local gauge symmetries play in determining the phase structure of a
  model, the nature of the different phases and of the quantum and
  thermal transitions.  Several lattice Abelian and non-Abelian gauge
  models have been considered, with the purpose of identifying the
  possible universality classes of the continuous transitions. In this
  paper we focus on systems characterized by an emerging discrete
  gauge symmetry, and in particular the ${\mathbb Z}_2$ gauge group.

\begin{figure}[tbp]
\includegraphics[width=0.9\columnwidth, clip]{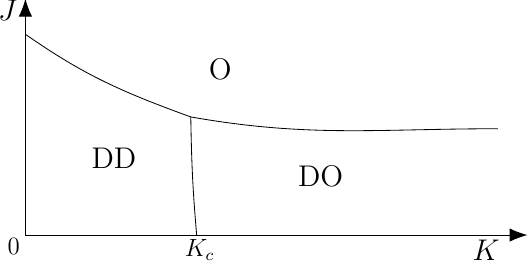}
\caption{Sketch of the phase diagram of the ${\mathbb Z}_2$-gauge
  $N$-vector model for $N\ge 2$, in the space of the Hamiltonian
  parameters $K$ and $J$, cf. Eq.~(\ref{ham}). For small $J$ there are
  two spin-disordered phases: a small-$K$ phase, in which both the
  spins and the gauge variables are disordered (indicated by DD), and
  a large-$K$ phase in which the ${\mathbb Z}_2$-gauge variables order
  (DO). In the large-$J$ phase both spins and gauge variables are
  ordered (indicated by O). The O($N$)$^*$ transition line is the one
  that separates the DO and O phase for sufficiently large $K$.  The
  three transition lines meet in a single point $(K_\star,J_\star)$;
  $(K_\star\approx 0.75,J_\star\approx 0.23)$ for small values of $N$,
  $N\lesssim 5$ say; see Ref.~\cite{BPV-24-z2gaugeN} for more
  details.}
\label{phadiaN}
\end{figure}

Lattice vector systems with ${\mathbb Z}_2$ gauge symmetry may develop
critical behaviors belonging to nonstandard $N$-vector universality
classes, in which the fundamental vector modes cannot be identified by
using gauge-invariant correlators; see, e.g., Refs.~\cite{Wegner-71,
  FS-79, Kogut-79, DH-81, SF-00, SSS-02, SM-02, SSSNH-02, CAP-08,
  HFS-11, IMH-12, PTV-18, Sachdev-19, WCWM-21, BP-23, ZZV-23,
  Senthil-23, BPV-24-decQ2, BPV-24-z2gaugeN}.  Such unconventional
O($N$) transitions occur for example in three-dimensional (3D)
${\mathbb Z}_2$-gauge $N$-vector models~\cite{BPV-24-z2gaugeN}, i.e.,
in lattice $N$-vector models in which the global ${\mathbb Z}_2$
symmetry is gauged, along the line that separates the spin disordered
phase from the spin ordered one, for sufficiently small values of the
gauge coupling, i.e., for large $K$ in the phase diagram sketched in
Fig.~\ref{phadiaN}.  These nonstandard O($N$) vector universality
classes, characterized by the symmetry breaking pattern
SO($N$)$\to$O($N-1$) and by the absence of a vector order parameter,
have been somehow distinguished by adding a star, i.e., by naming them
O($N$)$^*$ universality classes, see, e.~g.,
Ref.~\cite{Senthil-23}. Of course, the length-scale critical exponent
$\nu$ is the same in O($N$) and O($N$)$^*$ systems.

In our study we mostly discuss the O($N$)$^*$ transitions
  developed by the ${\mathbb Z}_2$-gauge $N$-vector models, which are
  relatively simple, but nontrivial, representatives of statistical
  systems undergoing this class of continuous transitions. However,
  the validity of our discussion extends to generic O($N$)$^*$
  transitions characterized by the absence of a local gauge-invariant
  vector order parameter.

In vector systems with global O($N$) symmetry, continuous transitions
are characterized by the spontaneous breaking of the O($N$) symmetry,
driven by the condensation of the $N$-component vector field.
However, in ${\mathbb Z}_2$-gauge $N$-vector models the correlations
of the local vector operator are trivial, as a consequence of the
${\mathbb Z}_2$-gauge symmetry. Therefore, the spontaneous breaking of
the O($N$) symmetry can only be observed by considering correlations
of composite gauge-invariant operators, the simplest one being an
operator that transforms as a spin-two tensor under O($N$)
transformations. At O($N)^*$ transitions, this operator, as well as
all gauge-invariant operators, have the same critical behavior as in
the conventional $N$-vector model without gauge invariance.  The
equivalence of the gauge-invariant correlations in O($N$) and O($N)^*$
transitions implies that gauge modes do not drive the critical
behavior. As a consequence, one should be able to describe these
transitions in terms of an effective Landau-Ginzburg-Wilson (LGW)
$\Phi^4$ field theory. The main issue here is the identification of
the correct fundamental field $\Phi$.  If one identifies $\Phi$ with a
coarse-grained gauge-invariant spin-2 order parameter, the LGW theory
is not able to properly describe the phenomenology of the O($N)^*$
transitions.  Indeed, O($N$)-symmetric transitions driven by the
condensation of a tensor spin-two field are characterized by a
different symmetry-breaking pattern, thus their nature differs from
that of the O($N$) vector transitions, see, e.g.,
Ref.~\cite{BPV-24-z2gaugeN}.

\begin{figure}[tbp]
\includegraphics[width=0.95\columnwidth, clip]{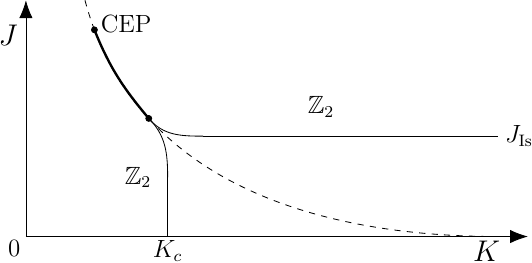}
\caption{Sketch of the phase diagram of the 3D ${\mathbb Z}_2$-gauge
  Higgs model. The dashed line is the self-dual line of the model, the
  thick line corresponds to first-order transitions along the
  self-dual line.  The two lines labelled ``${\mathbb Z}_2$" are
  related by duality and correspond to Ising continuous transitions.
  The transition lines meet at a multicritical point along the
  self-dual line, at~\cite{BPV-22-z2h,SSN-21,OKGR-23,XPK-24}
  $[K_\star=0.7525(1),J_\star\approx 0.22578(5)]$, where the
  multicritical behavior is controlled by a multicritical $XY$ fixed
  point~\cite{BPV-22-z2h,BPV-24-com,XPK-24}.  The Ising$^*$ transition
  line is the one ending at the Ising transition for $K=\infty$ and $J
  = J_{\rm Is}$.}
\label{phadiaz2}
\end{figure}

The critical behavior is even less conventional in the ${\mathbb
  Z}_2$-gauge Higgs model~\cite{Wegner-71,FS-79,Kogut-79}
(corresponding to a ${\mathbb Z}_2$-gauge $N$-vector model with
$N=1$). In this case there is no global ${\mathbb Z}_2$
  symmetry, but nonetheless, the transitions that occur for small
  gauge couplings, i.e., for large $K$, see the phase diagram sketched
  in Fig.~\ref{phadiaz2}, have the same universal features as Ising
  transitions, which are characterized by the breaking of a global
  ${\mathbb Z}_2$ symmetry in standard systems. We will refer to these
  transitions as Ising$^*$ transitions.  Because of duality, the same
  Ising behavior is observed on the small-$J$ line that starts at
  $J=0$ (${\mathbb Z}_2$ gauge model).  Also these transitions are
  sometime referred to as Ising$^*$ transitions, although the relation
  is obtained by the explicit use of duality; see, e.g.,
Refs.~\cite{TKPS-10,SWHSL-16,SSN-21,BPV-22-z2h}.

A natural question concerning the O($N$)$^*$ transitions is whether it
is possible to introduce a gauge fixing that allows the emergence of a
vector field that orders at the transition, so that it can be
identified as the order parameter for the spontaneous breaking of the
global O($N$) symmetry, as in standard $N$-vector models. We can ask
the same question for Ising$^*$ transitions. In this case the gauge
fixing should turn the local ${\mathbb Z}_2$ symmetry into a global
one, which is broken at the transition by the condensation of a scalar
order parameter. We mention that the search of (typically
  nonlocal) operators playing the role of ${\mathbb Z}_2$ order
  parameter at Ising$^*$ transitions has recently attracted much
  interest, see, e.g., Refs.~\cite{SSN-21,XPK-24,SSN-24}. One possible
  operator is the so-called Fredenhagen-Marcu order parameter
  \cite{FM-83}, that is an appropriate order parameter to characterize
  the Higgs phase in any lattice gauge theory. A different proposal,
  tailored for the ${\mathbb Z}_2$-gauge Higgs model, is presented in
  Ref.~\cite{SSN-21}. 

It is worth mentioning that there is another class of continuous
transitions, in which critical vector modes cannot be observed by
using gauge-invariant correlators.  We refer to the transitions
between the Coulomb and Higgs phases in noncompact lattice Abelian
Higgs (AH) models, in which a complex $N$-vector field is minimally
coupled with a noncompact U(1) gauge field, see, e.~g.,
Refs.~\cite{HLM-74,KK-85,KK-86,BN-87,MV-04,IZMHS-19,
  BPV-21-ncAH,BPV-22,BPV-23-chgf,BPV-24-ncAH}.  An effective
description of these {\em charged} transitions is provided by the AH
field theory, in which a vector and a gauge field are the fundamental
variables that drive the critical behavior.  However, in lattice
models, the fundamental vector field does not show critical
correlations, because of gauge invariance. The puzzle was solved in
Refs.~\cite{BPV-23-chgf,BPV-24-ncAH}, where it was shown that the
correct order parameter for these transitions is a nonlocal
gauge-invariant charged vector operator~\cite{KK-85, KK-86,
  BPV-23-chgf, BPV-24-ncAH}.  Equivalently, a local critical vector
field is obtained by using the Lorenz gauge fixing~\cite{BPV-23-gf}.

In this paper, we show that, as in noncompact AH models, it is
possible to identify a vector order parameter for O($N$)$^*$
transitions, by means of an appropriate gauge-fixing
procedure. However, the approach needed here---we name it stochastic
gauge fixing--- is more complicated than the Lorenz gauge fixing
working for noncompact lattice AH models.  In this approach the
gauge-fixed vector correlations show the universal critical behavior
expected at transitions belonging to the standard O($N$) vector
universality class.  Analogously, in the ${\mathbb Z}_2$-gauge Higgs
model, we are able to observe the universal critical correlations of
an emerging ${\mathbb Z}_2$ order parameter.  This shows that the
O($N$)$^*$/Ising$^*$ universality class is equivalent to the more
standard O($N$)/Ising vector universality class, characterized by the
condensation of a vector/scalar order parameter.  To validate the
approach, we present numerical finite-size scaling (FSS)
analyses of Monte Carlo (MC) data for $N=1$ and 3.  Some results for
$N=2$ were already reported in Ref.~\cite{BPV-24-z2gaugeN}.

The paper is organized as follows. In Sec.~\ref{model} we introduce
the ${\mathbb Z}_2$-gauge $N$-vector models and the ${\mathbb
  Z}_2$-gauge Higgs model corresponding to $N=1$, and summarize the
main features of their phase diagrams. In Sec.~\ref{gaufix} we define
the stochastic gauge fixing, which allows us to uncover the critical
vector correlations along the O($N$)$^*$ transition lines, and we
discuss the relation between the stochastic gauge fixing scheme and
random-bond Ising systems.  In Sec.~\ref{numres} we report a numerical
study of the O(3)$^*$ transitions and Ising$^*$ transitions. In
particular, we show that vector correlation functions are critical, in
the stochastic gauge fixing scheme.  Finally, in Sec.~\ref{conclu} we
summarize and draw our conclusions.

\section{The ${\mathbb Z}_2$-gauge $N$-vector model}
\label{model}

The lattice ${\mathbb Z}_2$-gauge $N$-vector model is defined on a 3D
cubic lattice. Its Hamiltonian is
\begin{eqnarray}
&& H(J,K) = H_{\bm s}(J) + H_\sigma(K), \label{ham}\\
&& H_{\bm s}(J) = - J N \sum_{{\bm x},\mu} \sigma_{{\bm x},\mu} \, {\bm s}_{\bm
    x} \cdot {\bm s}_{{\bm x}+\hat{\mu}}, \label{hs}\\
&& H_\sigma(K) =  - K \sum_{{\bm
      x},\mu>\nu}
   \sigma_{{\bm
      x},\mu} \,\sigma_{{\bm x}+\hat{\mu},\nu} \,\sigma_{{\bm
      x}+\hat{\nu},\mu} \,\sigma_{{\bm x},\nu},
\label{hsi}
\end{eqnarray}
where the site variables ${\bm s}_{\bm x}$ are unit-length
$N$-component real vectors, and the bond variables $\sigma_{{\bm
    x},\mu}$ ($\sigma_{{\bm x},\mu}$ is associated with the bond
starting from site ${\bm x}$ in the positive $\mu$ direction,
$\mu=1,2,3$) take the values $\pm 1$. The Hamiltonian parameter $K$
plays the role of inverse gauge coupling, therefore the $K\to\infty$
limit corresponds to the small gauge-coupling limit.  By measuring
energies in units of the temperature $T$, we can formally set $T=1$
and write the partition function as
\begin{equation}
Z(J,K)=\sum_{\{{\bm s},\sigma\}} e^{-H(J,K)}.
\label{partfunc}
\end{equation}
For $N=1$ the spin variables take the integer values $s_{\bm x}=\pm
1$, and the model corresponds to the so-called ${\mathbb Z}_2$-gauge
Higgs model~\cite{Wegner-71,FS-79,Kogut-79}.

The Hamiltonian (\ref{ham}) is invariant under global SO($N$)
transformations ${\bm s}_{\bm x} \to V {\bm s}_{\bm x}$ with $V\in
\mathrm{SO}(N)$, and local ${\mathbb Z}_2$ gauge transformations,
\begin{eqnarray}
{\bm s}_{\bm x}\to w_{\bm x} {\bm s}_{\bm x},\quad
\sigma_{{\bm x},\nu}\to 
w_{\bm x} \sigma_{{\bm x},\nu} w_{{\bm x}+\hat{\nu}},\quad w_{\bm x}=\pm 1.
\label{gautra}
\end{eqnarray}
From the point of view of the symmetries, the model can be interpreted
as an $N$-vector model, which is O($N$) = ${\mathbb Z}_2\otimes
\mathrm{SO}(N)$ symmetric, in which the ${\mathbb Z}_2$ symmetry is
gauged, i.e., becomes local.  Due to the ${\mathbb Z}_2$ gauge
invariance, the correlation function of the vector variables ${\bm
  s}_{x}$,
\begin{eqnarray}
G_s({\bm x},{\bm y}) = \langle {\bm s}_{\bm x} \cdot {\bm s}_{\bm y} \rangle,
\label{vectcorr}
\end{eqnarray}
trivially vanishes for ${\bm x}\neq {\bm y}$ and any Hamiltonian
parameter $K$ and $J$.  For the same reason, any correlation function
of local operators defined as products of an odd number of spin
variables (such as a spin-3 local operator) vanishes as well. These
correlations, therefore, cannot characterize the O($N$)$^*$
transitions occurring for large $K$, see
Figs.~\ref{phadiaN}~and~\ref{phadiaz2}.  For $N\ge 2$, the spontaneous
breaking of the global O($N$) symmetry is instead signaled by the
condensation of the gauge-invariant bilinear spin-two operator
\begin{eqnarray}
  Q^{ab}_{\bm x} = s_{\bm x}^a s_{\bm x}^b - {1\over N}\delta^{ab}.
  \label{qab}
\end{eqnarray}

The ${\mathbb Z}_2$-gauge $N$-vector model is a paradigmatic model
relevant for transitions in nematic liquid crystal, see, e.g.,
Refs.~\cite{LRT-93,KNNSWS-15}, and for systems with fractionalized
quantum numbers, see, e.g., Refs.~\cite{SSS-02,SM-02}. It shows
different phases characterized by the spontaneous breaking of the
global SO($N$) symmetry and by the different topological properties of
the ${\mathbb Z}_2$-gauge correlations, see, e.g.,
Refs.~\cite{FS-79,Sachdev-19,BPV-24-z2gaugeN}.  Its phase diagram for
$N\ge 2$ is sketched in Fig.~\ref{phadiaN}.  Two spin-disordered
phases are present for small $J$: a small-$K$ phase, in which both
spin and ${\mathbb Z}_2$-gauge variables are disordered (DD), and a
large-$K$ phase in which the ${\mathbb Z}_2$-gauge variables order
(DO).  For large $J$ there is a single phase in which both spins and
gauge variables order (O).  These phases are separated by three
transition lines.  The transitions along the DD-DO line, departing
from the $J=0$ axis at ~\cite{BPV-20-hcAH,FXL-18} $K_c(J=0) =
0.761413292(11)$, are continuous, at least for small enough values of
$J$, and belong to the ${\mathbb Z}_2$-gauge universality
class~\cite{Wegner-71,FS-79,Sachdev-19} for any $N$. Instead, the main
features of the DD-O and DO-O transitions crucially depend on $N$ and
are discussed in Ref.~\cite{BPV-24-z2gaugeN}.  Here, for the
convenience of the reader, we summarize the main characteristics of
the DO-O transition line, which is the one relevant for the present
study.  The transitions along the DD-O line are more
conventional. They can be associated with an effective LGW theory in
which the fundamental field is obtained by coarse-graining the order
parameter defined in Eq.~\eqref{qab}. The three transition lines meet
in one point, see Fig.~\ref{phadiaN}, located at $[K_\star\approx
  0.75,J_\star\approx 0.23]$ for sufficiently small values of $N$,
$N\lesssim 5$ say.\footnote{Since $K_c(J) \approx K_c(J=0) - N
J^4$~\cite{BPV-24-z2gaugeN} for small $J$ and $J_\star$ is only
slightly larger than $J_c(K=\infty)= J_{{\rm O}(N)} \lesssim 0.25$
($J_{{\rm O}(N)}$ is the critical point of the standard $N$-vector
model), we expect $K_\star \approx 0.761 - 0.003 N$ for sufficiently
small values of $N$.}

The transitions along the DO-O line are expected to belong to the
O($N$)$^*$ universality class. This is due to the stability of the
O($N$) vector transition occurring for $K\to\infty$ against gauge
fluctuations.  Indeed, for $K\to\infty$ the plaquette term in the
Hamiltonian (\ref{ham}) converges to 1. In infinite volume, we can
just set $\sigma_{{\bm x},\mu} = 1$ modulo gauge transformations,
obtaining the partition function of the standard $N$-vector model.
Therefore, for $K\to\infty$ the model (\ref{ham}) undergoes a
continuous transition at a finite $J_{c}(K=\infty)$ belonging to the
O($N$) vector universality class (estimates of $J_{c}(K=\infty)$,
i.e., of the critical point of the standard $N$-vector models, can be
found in
Refs.~\cite{Hasenbusch-19,DBN-05,BFMM-96,Hasenbusch-22,DPV-15,BC-97,CPRV-96}).
The O($N$) transition is expected to be stable against small gauge
fluctuations, for sufficiently large but finite values of
$K$~\cite{BPV-24-z2gaugeN}, due to the discrete nature of the gauge
variables, whose fluctuations are suppressed in the large-$K$
topologically ordered phase. Therefore, the continuous DO-O
transitions are expected to belong to the O($N$)$^*$ universality
class. Note that gauge fluctuations are instead relevant in models
with continuous Abelian and non-Abelian gauge symmetries.  In that
case gauge interactions destabilize the vector critical behavior,
leading to transitions of different nature, see, e.g.,
Refs.~\cite{BPV-19,BPV-22,BPV-23-chgf,BPV-23-mppo,BPV-24-ncAH}.  If
the DO-O transitions belong to the ${\rm O}(N)^*$ universality class,
we expect the presence of critical vector modes. But, because of gauge
invariance, they cannot be directly identified; for instance, the
correlation function $G_s$ defined in Eq.~(\ref{vectcorr}) is not
critical. They however emerge if an appropriate gauge-fixing procedure
is considered, as we discuss below.

The phase diagram for $N=1$ is shown in Fig.~\ref{phadiaz2}, see,
e.g., Refs.~\cite{BPV-22,SSN-21,TKPS-10}.  Unlike the multicomponent
$N\ge 2$ models, only two phases are present, separated by two
${\mathbb Z}_2$ lines that are related by duality~\cite{BDI-74} and
correspond to Ising continuous transitions.  They end at~\cite{FXL-18}
$[J = J_{\rm Is} = 0.221654626(5), K=\infty]$ and at $[J =0,K = K_c =
  0.761413292(11)]$ and meet in a multicritical $XY$ point along the
self-dual line, located at $[K_\star=0.7525(1),J_\star\approx
  0.22578(5)]$, see, e.g..
Refs.~\cite{TKPS-10,SSN-21,BPV-22,OKGR-23,XPK-24}.  The endpoint of
the first-order transition line, at $[K\approx 0.688,J\approx 0.258]$,
is expected to be an Ising critical endpoint.  Transitions along the
large-$K$ transition line, running almost parallel to the $K$ axis,
belong to the Ising$^*$ universality class. This is analogous to what
occurs along the DO-O line for $N\ge 2$, where transitions belong to
the O($N$)$^*$ universality class. However, for $N=1$ there are no
gauge-invariant spin correlations.  Critical scalar correlations
emerge only after implementing the stochastic gauge fixing that we
will outline in the next section.
  
\section{Stochastic gauge fixing}
\label{gaufix}

\subsection{General considerations on the standard gauge-fixing approach}
\label{sec3.1}

In lattice models with noncompact Abelian variables, it is relatively
easy to define a consistent gauge fixing procedure, see, e.g.,
Refs.~\cite{BPV-23-gf,BPV-23-chgf,BPV-24-ncAH} and references therein.
In these models, indeed, the lattice gauge variables take values in
$\mathbb{R}$, and gauge fixing can be introduced just like in
continuum theories; in particular, gauge fixing procedures defined by
linear functionals (like the Lorenz gauge) are particularly
straightforward to implement. In models with discrete or compact bond
variables, instead, due to the intrinsic nonlinear structure of the
gauge fixing, several additional problems arise, even for Abelian
gauge groups. We discuss here this issue for the ${\mathbb Z}_2$ gauge
group but the discussion can easily be extended to any Abelian gauge
group.

Since our aim is to use gauge fixing to expose the criticality of
  non gauge invariant quantities (and in particular of the would be
  order parameter), two different problems must be tackled. The first
  problem is the definition and consistency of the gauge fixing
  procedure, which means that we must show that expectation values of
  gauge invariant quantities are the same in the original theory and
  in the gauge fixed one. The second problem concerns the existence of
  a nontrivial critical behavior of the gauge variant modes in the
  gauge fixed theory, which is by no means guaranteed. Examples of
  gauge fixing procedures which are not useful in studying the
  critical properties of noncompact Abelian gauge models are discussed
  in Refs.~\cite{BPV-23-gf,BPV-23-chgf}.

Let us start investigating the first of these two problems.  In
general, a gauge fixing is defined by a local function of the gauge
fields $F_{\bm x}(\sigma)$, and by the requirement that $F_{\bm
  x}(\sigma) = 0$ for all lattice point $x$. To avoid the problem of
the Gribov copies~\cite{Gribov:1977wm, Singer:1978dk}, we assume that
the gauge fixing is complete, i.e., that for each configuration
$\{\sigma_{{\bm x},\mu}\}$ there is a unique configuration
$\{\sigma_{{\bm x},\mu}' \}$, related to $\{\sigma_{{\bm x},\mu}\}$ by
a gauge transformation, such that $F_{\bm x}(\sigma') = 0$. Although
the relation between $\{\sigma_{{\bm x},\mu}\}$ and $\{\sigma_{{\bm
    x},\mu}' \}$ is unique, the gauge transformation that relates the
two configurations is not. However, if $w^{(a)}_{\bm x}$ and
$w^{(b)}_{\bm x}$ are two gauge transformations that both relate
$\{\sigma_{{\bm x},\mu}\}$ with $\{\sigma_{{\bm x},\mu}' \}$, it is
trivial to show that $w^{(a)}_{\bm x} = c \,w^{(b)}_{\bm x}$, where $c
= \pm 1$. This result implies that the completeness of the gauge
fixing does not imply a unique definition of the gauge-fixed spin.  In
the gauge fixing procedure that relates $\{\sigma_{{\bm x},\mu}\}$
with $\{\sigma_{{\bm x},\mu}' \}$, the new spin $s_{\bm x}'$ is only
defined up to a sign (the constant $c$ defined above). This is not a
limitation, if we only consider correlation functions with an even
number of spins, for instance, the two-point function. Moreover, the
previous result implies
\begin{equation}\label{eq:FP}
   \sum_{w} \frac{1}{2}\prod_{\bm x} \delta[F_{\bm x}(\sigma')]
      = 1, \qquad 
    \sigma'_{{\bm x},\mu} = 
     w_{\bm x}  \sigma_{{\bm x},\mu} w_{{\bm x} + \hat{\mu}},
\end{equation}
independently of $\sigma_{{\bm x},\mu}$, which is just the
  formalization of the previous statement that two different lattice
  gauge fields always correspond to the gauge fixed $\{\sigma_{{\bm
      x},\mu}'\}$ gauge field.  This is just a discrete version of the
  standard Faddeev-Popov procedure \cite{FP-67} used to properly
  define gauge fixing in continuum field theories, and in the most
  general case a weight depending on $\sigma_{{\bm x},\mu}'$ appears
  instead of the constant factor $1/2$ in Eq.~\eqref{eq:FP}, which is
  typically rewritten by using auxilliary field variables (the so
  called Faddeev-Popov ghosts). By inserting the identity
  Eq.~\eqref{eq:FP} in the sum in Eq.~\eqref{partfunc} defining the
  partition function, it immediately follows that correlation
  functions of gauge-invariant operators computed in the gauge-fixed
  theory and in the original theory are the same.  We have thus shown
  that in a $\mathbb{Z}_2$ lattice gauge theory any complete gauge
  fixing defined by a local functional $F_{\bm x}(\sigma)$ can be
  used, without having to worry about complications related to ghosts
  fields.

A standard way of fixing the gauge consists in setting the bond
variables on a maximal lattice tree equal to the
identity~\cite{Creutz-77, OS-78}. A particular case is the axial
gauge, in which all bonds in a given lattice direction are set equal
to the identity, paying attention to the boundary conditions and
adding some additional constraints on the boundaries.  Another
possibility that we considered is the gauge fixing obtained by using
the gauge function
\begin{equation}
  f_{\bm x} \equiv -1 + \prod_{\mu} \sigma_{{\bm x}-
    \hat{\mu},\mu}\sigma_{\bm{x},\mu},
  \label{lorequ}
\end{equation}
which somehow generalizes the Lorenz gauge of noncompact Abelian gauge
theories. If we define $F_{\bm x} = f_{\bm x}$ on the whole lattice,
then the gauge fixing is not complete.  It is however possible to make
it complete by changing the gauge-fixing function on a lattice
boundary (we do not report details, as this approach turns out to
fail). 

These types of gauge fixings do not however allow us to identify the
critical vector modes that characterize the O($N$) vector universality
class. For instance, the gauge-fixed correlation function $\langle
{\bm s}_{\bm x}'\cdot {\bm s}_{\bm y}'\rangle$ [the fields ${\bm
    s}'_{\bm x}$ are those obtained imposing the gauge-fixing
  constraint (\ref{lorequ})] is not critical along the O($N$)$^*$
transition line.  The absence of criticality can be traced back to the
fact that the gauge-fixing procedure is strictly nonlocal. As a
consequence a local change of the initial configuration $\sigma_{{\bm
    x},\mu}$ may give rise to a nonlocal change of the gauge-fixed
configuration $\sigma_{{\bm x},\mu}'$, which prevents the spins
  $s_{\bm x}$ from acquiring a nonvanishing polarization. This
phenomenon is easy to understand in the axial gauge defined by
  $\sigma_{{\bm x},3}'=1$, but the same occurs when using the
gauge-fixing constraint (\ref{lorequ}).  If the original (i.e.,
  not gauge fixed) configuration is $\sigma_{{\bm x},\mu} = 1$ on the
  whole lattice, then we also have $\sigma_{{\bm x},\mu}'= 1$; the
  gauge fixed spins $s'_{\bm x}$ thus behave as in the O($N$) vector
  model and, in particular, they are ferromagnetically ordered for
  large $J$. If however the original configuration has a single single
  bond misaligned in the axial direction in the bulk of the lattice,
  i.~e. $\sigma_{{\bm y},3}=-1$ for a single ${\bm y}$ value in the
  bulk, in the gauge fixed configuration $\sigma_{{\bm x},\mu}'=-1$ on
  a number of links of order $L^3$. As a consequence $s'_{\bm x}$ does
  not order ferromagnetically for large $J$. Thus, even when $K$ is
  very large, typical configurations $\{ s'_{\bm x}\}$ are not related
  with the typical configurations of the O($N$) vector
  model. Therefore, $\langle {\bm s}_{\bm x}'\cdot {\bm s}_{\bm
    y}'\rangle$ does not show any ferromagnetic order for large $J$.

\subsection{A new approach} \label{sec3.2}

In this work we pursue a different approach that allows us to uncover
critical gauge-dependent order-parameter correlations.  The basic idea
is to average non-gauge invariant quantities over all possible gauge
transformations with a properly chosen, nongauge invariant, weight. We
introduce ${\mathbb Z}_2$ fields $w_{\bm x}=\pm 1$ defined on the
lattice sites [associated with the local gauge transformations,
  see Eq.~\eqref{gautra}], and an ancillary Hamiltonian $H_w$ that
generally depends on $w_{\bm x}$, ${\bm s}_{\bm x}$, and $\sigma_{{\bm
    x},\mu}$. If $A({\bm s}_{\bm x},\sigma_{{\bm x},\mu})$ is a
function of the fields, we define its weighted average over the gauge
transformations as
\begin{eqnarray} 
[A({\bm s}_{\bm x},\sigma_{{\bm x},\mu})] = 
{\sum_{\{w\}} A(\hat{\bm s}_{\bm x},
  \hat\sigma_{{\bm x},\mu}) e^{-H_w} \over
  \sum_{\{w\}}  e^{-H_w} },
\label{gfav}
\end{eqnarray}
where $\hat{\bm s}_{\bm x}$ and $\hat\sigma_{{\bm x},\mu}$ are defined as 
\begin{eqnarray}
  \hat{\bm s}_{\bm x} = w_{\bm x}{\bm s}_{\bm x}, \qquad
  \hat\sigma_{{\bm x},\mu} = w_{\bm x}\sigma_{{\bm x},\mu} w_{{\bm
      x}+\hat{\mu}},
\label{gautra2}
\end{eqnarray}
and correspond to the fields obtained by performing a gauge
transformation with gauge function $w_{\bm x}$.  The average
$[A(\bm s_{\bm x},\sigma_{{\bm x},\mu})]$ is then averaged
over the fields ${\bm s}_{\bm x}$ and $\sigma_{{\bm x},\mu}$ using the
original Hamiltonian (\ref{ham}), i.e.
\begin{eqnarray}
  \langle [A({\bm s}_{\bm x},\sigma_{{\bm x},\mu})] \rangle =
{\sum_{\{{\bm s},\sigma\}} [A(\bm s_{\bm x},
  \sigma_{{\bm x},\mu})] e^{-H} \over
  \sum_{\{{\bm s},\sigma\}}  e^{-H} }.
\label{finav}
\end{eqnarray}
Gauge-invariant observables are invariant under the procedure.
Indeed, if $A({\bm s}_{\bm x},\sigma_{{\bm x},\mu})$ is a
gauge-invariant observable, then
\begin{equation}
A(\hat{\bm s}_{\bm x},\hat\sigma_{{\bm x},\mu}) =  
A({\bm s}_{\bm x},\sigma_{{\bm x},\mu}) =
[A({\bm s}_{\bm x},\sigma_{{\bm x},\mu})].
\label{gauiave}
\end{equation}
In this approach, we define a vector correlation function as
\begin{equation}
G_V({\bm x},{\bm y}) = \langle [{\bm s}_{\bm x} \cdot {\bm
    s}_{\bm y}] \rangle.
\label{gvgf}
\end{equation}
A crucial point in the procedure is the choice of the Hamiltonian $H_w$. To
unveil O($N$) vector correlations, we would like to work in a gauge which
maximizes the number of bonds with $\hat\sigma_{{\bm x},\mu}= 1$. Indeed,
this implies that the Hamiltonian for the gauge-transformed fields
$\hat{\bm s}_{\bm x}$ is almost ferromagnetic. Therefore, these fields
display the same critical behavior as vector fields in the O($N$) model.

With this idea in mind, we consider 
\begin{equation}
  H_w(\gamma)=- \gamma \sum_{{\bm x},\mu} \hat\sigma_{{\bm x},\mu} = -
  \gamma \sum_{{\bm x},\mu} w_{\bm x} \sigma_{{\bm x},\mu} w_{{\bm
      x}+\hat\mu},
\label{Hw}
\end{equation}
where $\gamma$ is a positive number that should be large enough---this
point will be discussed in detail below---to ensure that the minima of
$H_w(\gamma)$ dominate in the average over the gauge transformations.

This procedure, which we call stochastic gauge fixing, mimics what is
done in random systems with quenched disorder, for instance in spin
glasses.  The variables ${\bm s}_{\bm x}$ and $\sigma_{{\bm x},\mu}$
are the disorder variables and $e^{-H}/Z$ represents the disorder
distribution, while $w_{\bm x}$ are the system variables that are
distributed with Gibbs weight $e^{-H_w}/Z_w$ at fixed disorder. In the
language of disordered systems, the average $[\cdot ]$ therefore
represents the thermal average at fixed disorder, while $\langle
\cdot\rangle$ is the average over the different disorder
realizations.\footnote{To avoid confusion, note that symbols $[\cdot
]$ and $\langle \cdot\rangle$ have typically the opposite meaning in
the random-system literature: the former represents the disorder
average and the latter the thermal average.}  This analogy allows us
to use the wealth of available result for quenched random systems. In
particular, the present procedure is thermodynamically consistent and,
when the low-temperature (large $\gamma$) phase is not a spin-glass
phase, it admits a local field-theory representation.  Thus, we can
apply the standard renormalization-group (RG) machinery to
correlations computed in the gauge-fixed theory.

The resulting model with the added variables $w_{\bm x}$ is a quenched
random-bond Ising model~\cite{Harris-74,EA-75} with a particular
choice of bond distribution.  Quenched random-bond Ising models have
various phases---disordered, ferromagnetic, and glassy
phases---depending on the temperature (whose role is played here by
$1/\gamma$), the amount of randomness of the bond distribution, and
its spatial correlations, see, e.g.,
Refs.~\cite{Harris-74,EA-75,Nishimori-81,LH-88,PC-99,Betal-00,Nishimori-book,
  KKY-06,HPPV-07,HPV-08,CPV-11,Janus-13,LPP-16,CP-19}.  In particular,
we expect the present model to undergo a quenched transition for
$\gamma = \gamma_c(J,K)$. The transition separates a disordered phase
for $\gamma < \gamma_c(J,K)$ from a large-$\gamma$ phase, which a
priori can be ferromagnetic or glassy.  As we shall discuss, if $J$
and $K$ belong to the DO-O transition line, the large-$\gamma$ phase
is ferromagnetic and the minimum configurations are essentially unique
modulo global symmetries.  Thus, the long-distance behavior of the
variables $w_{\bm x}$ is the same for all $\gamma > \gamma_c(J,K)$:
The variables $w_{\bm x}$ simply make uncorrelated short-range
fluctuations around the minimum configurations obtained for $\gamma
\to \infty$. It is thus natural to conjecture that $\gamma$ is an
irrelevant parameter, i.e., the critical behavior of the gauge-fixed
quantities is the same for any $\gamma>\gamma_c(J,K)$ along the DO-O
transition line.  In RG language, $1/\gamma$ represents an irrelevant
perturbation of the $\gamma=\infty$ fixed point.

We expect the irrelevance of $\gamma$ to be a general feature of the
stochastic gauge fixing, which holds for any $N$---the numerical data
we will present confirm this conjecture. Thus, for numerical
convenience, we will always implement the gauge-fixing procedure using
a finite value of $\gamma$.  It is important not to confuse this type
of gauge fixing with the more standard soft gauge fixing that is
obtained by adding to the Hamiltonian a term of the form $\lambda
F_{\bm x}^2$ involving a further parameter $\lambda$ (where the
equation $F_{\bm x}=0$ represents the hard gauge fixing), which in the
language of random systems represents an annealed average over the
gauge fixing. For example, in the noncompact AH model $\lambda^{-1}$ is
known to be a relevant perturbation of the Lorenz gauge-fixed theory,
see Refs.~\cite{BPV-23-gf, BPV-23-chgf}.

It is worth noting that the global theory including the quenched
stochastic gauge fixing is invariant under an extended set of local
transformations with ${\mathbb Z}_2$-gauge parameter $v_{\bm x}=\pm 1$
given by
\begin{eqnarray}\label{gaugetr-2}
{\bm s}_{\bm x} \to v_{\bm x} {\bm s}_{\bm x}, \;\;
\sigma_{{\bm x},\mu}  \to
v_{\bm x} \sigma_{{\bm x},\mu} v_{{\bm x}+\hat{\mu}}, \;\;
w_{\bm x} \to v_{\bm x} w_{\bm x}.
\end{eqnarray}
Therefore, only observables that are invariant under this set of
transformations are relevant, such as $\hat{s}_{\bm x}$ and
$\hat{\sigma}_{{\bm x},\mu}$. Note that this local extended
  symmetry prevents the $w_{\bm x}$ variables from acquiring a
  nonvanishing expectation value, and thus the presence of a standard
  ferromagnetic phase in this model of spin glass.

We remark that there is much freedom in the choice of the ancillary
Hamiltonian $H_w$. The Hamiltonian~\eqref{Hw} is quite appealing,
since it provides a direct connection with quenched models already
well investigated and is particularly simple.  However, other choices
may work as well.  It is important to note that the critical behavior
of the emerging vector critical correlators is expected to be
universal, i.e., independent of the ancillary Hamiltonian $H_w$,
provided that $H_w$ has been properly chosen to make the spin-spin
interactions ferromagnetic. This is essentially due to the fact that
the critical behavior of all gauge invariant quantities---for
instance, the spin-two operator $Q_{\bm x}$ or the cumulants of the
gauge-invariant energy---is independent of $H_w$: they all behave as
in the $N$-vector/Ising model. Therefore, along the DO-O line (or
along the corresponding transition line for $N=1$), $G_V({\bm x} -
{\bm y})$ should also behave as in the $N$-vector/Ising model, if it
is critical.

It is interesting to verify the general ideas of the approach in the
limit $K\to\infty$. In this limit we have
\begin{equation}
\sigma_{{\bm x},\mu} \,\sigma_{{\bm
    x}+\hat{\mu},\nu} \,\sigma_{{\bm x}+\hat{\nu},\mu} \,\sigma_{{\bm
    x},\nu} = 1,
\label{Plaq1}
\end{equation}
which implies $\sigma_{x,\mu} = \rho_{\bm x}
\rho_{{ \bm x}+\hat{\mu}}$, with $\rho_{\bm x}=\pm 1$, 
in the thermodynamic limit. It follows
\begin{eqnarray}
    H_J &=& - N J \sum_{{\bm x} ,\mu} (\rho_{\bm x} {\bm s}_{\bm x}) \cdot 
      (\rho_{{\bm x} + \hat\mu} {\bm s}_{{\bm x} + \hat\mu}), \nonumber \\
    H_w &=& - \gamma \sum_{{\bm x} ,\mu} (\rho_{\bm x} w_{\bm x}) \cdot 
      (\rho_{{\bm x} + \hat\mu} w_{{\bm x} + \hat\mu}).
\end{eqnarray}
Thus, if we redefine $w_{\bm x}' = \rho_{\bm x} w_{\bm x}$ and ${\bm
  s}_{\bm x}' = \rho_{\bm x} {\bm s}_{\bm x}$ [which are
    invariant under the local transformations of
    Eq.~\eqref{gaugetr-2}], the partition function factorizes.
Moreover, since $\rho_{\bm x}^2 = 1$, $G_V({\bm x})$ can be written as
\begin{equation}
   G_V({\bm x}-{\bm y}) = \langle {\bm s}'_{\bm x} \cdot {\bm s}'_{\bm y}
           \rangle_{{\rm O}(N),J} 
           \langle w'_{\bm x} \cdot w_{\bm y}'
           \rangle_{{\rm Is},\gamma},
\label{GV-factorized}
\end{equation}
where the two averages are  performed in the standard O($N$) and Ising 
model, respectively. For any $\gamma > \gamma_{c,\rm Is}$, the Ising system 
is magnetized and therefore $G_V({\bm x})$ has the same critical 
behavior as in the $N$-vector model, confirming the irrelevance of $\gamma$.
On the other hand, for $\gamma< \gamma_{c,\rm Is}$, $G_V({\bm x})$ 
is always disordered, irrespective of $J$. 

Note that Eq.~(\ref{Plaq1}) also holds for $J\to \infty$ for any value
of $K$. Thus, also in this case the partition function factorizes. The
gauge fields $w_{\bm x}$ behave as in the Ising model and so does the
correlation function $G_V({\bm x}-{\bm y})$ as a consequence of
Eq.~(\ref{GV-factorized}).

We finally mention that the above ideas can be straightforwardly
extended to lattice gauge models with other discrete groups or with
continuous gauge U(1) variables, thereby allowing one to uncover
order-parameter correlations that are not accessible using
gauge-invariant operators.

\section{Numerical results}
\label{numres}

In this section we discuss the critical behavior of the O($3$)$^*$ and
Ising$^*$ transitions in the ${\mathbb Z}_2$-gauge model for $N=3$ and
$N=1$, respectively.  In particular, we study the correlation function
$G_V({\bm x})$ defined in Eq.~(\ref{gvgf}) using the stochastic gauge
fixing approach. The numerical analysis of the data shows that
$G_V({\bm x})$ behaves as the vector correlation function in standard
$N$-vector models.  This result provides the last piece of evidence
for the identification of the transitions as O($N$) (or Ising)
transitions driven by the condensation of a vector, but
gauge-dependent, order parameter.

\subsection{Monte Carlo simulations}
\label{mcsim}

We perform simulations of the ${\mathbb Z}_2$-gauge model for $N=1$
and $N=3$ applying the stochastic gauge fixing.  In both cases, we
perform runs at fixed $K$ varying $J$ around the critical point
$J=J_c$, where the model undergoes an O(3)$^*$ or Ising$^*$ transition
for $N=3$ and $N=1$, respectively. We choose $K=1$, which is larger
than $K_\star \approx 0.75$, the value of the meeting point of the
three transition lines, see Figs.~\ref{phadiaN} and \ref{phadiaz2}.
The parameter $\gamma$ of the ancillary Hamiltonian $H_w$ should be
large enough to guarantee that the ancillary quenched system is in the
ferromagnetic phase, as discussed in Sec.~\ref{gaufix}.  For $N=3$ we
find $\gamma_c(J_c,K=1)\approx 0.23$, and we thus fix
$\gamma=0.3$. Also for $N=1$ the ancillary quenched system is in the
ferromagnetic phase when using $\gamma=0.3$ at $K=1$ and $J=J_c$, so
we use $\gamma=0.3$ for both values of $N$.

MC simulations are performed as in systems with quenched disorder,
see, e.g.,
Refs.~\cite{Harris-74,EA-75,Nishimori-81,LH-88,PC-99,Betal-00,Nishimori-book,
  KKY-06,HPPV-07,HPV-08,CPV-11,Janus-13,LPP-16,CP-19}.  We simulate
the model with Hamiltonian $H$ and every $N_s$ update sweeps we
compute the quenched averages $[A]$ over the gauge-fixing variables
for fixed values of ${\bm s}_{\bm x}$ and $\sigma_{{\bm x},\mu}$ (at
fixed disorder in the language of random systems). Simulations are
performed by using standard local Metropolis updates. For $N=3$, we
also perform microcanonical updates of the ${\bm s}_{\bm x}$
variables. Quenched averages are computed using $10^4$ complete update
sweeps of the $w_{\bm x}$ variables, at fixed $\sigma_{{\bm x},\mu}$
and ${\bm s}_{\bm x}$.

Before presenting our results, we report the values of 
the critical exponents for the 3D O(3) (Heisenberg) and Ising 
universality classes, that are used in our analyses.
These exponents  are known with great accuracy, see, e.g., 
Refs.~\cite{Hasenbusch-20-o3,Chester-etal-20-o3,KP-17,HV-11,
  CHPRV-02,GZ-98,PV-02} for $N=3$ and
Refs.~\cite{PV-02,GZ-98,CPRV-02,Hasenbusch-10,KPSV-16,
  KP-17,FXL-18,Hasenbusch-21,Hasenbusch-23} for $N=1$.
For the 3D Heisenberg universality class
accurate estimates are reported in 
Ref.~\cite{Hasenbusch-20-o3}:
\begin{equation}
  \begin{aligned}
  \nu_H &= 0.71164(10),\; \\ \eta_H &=0.03784(5),\; \\
  \omega_H &=0.759(2).
  \end{aligned}
  \label{Heiexp}
\end{equation}
Here $\nu_H$ is the correlation-length exponent, $\eta_H$ parametrizes
the behavior of the critical two-point function of the vector field,
and $\omega_H$ is the leading scaling-correction exponent.  In our FSS
analyses we also need the RG dimension $Y_Q$ of the spin-two composite
operator $Q_{\bm x}$ for the Heisenberg universality class
~\cite{HV-11,Hasenbusch-23,CPV-03,CPV-02}:
\begin{equation}
  Y_{QH} = 1.2094(3).
  \label{YQH}
\end{equation}
For the 3D Ising universality class, we report~\cite{KPSV-16}
\begin{equation}
\begin{aligned}
\nu_I &= 0.629971(4),\;\\
\eta_I &= 0.036298(2),\;\\
\omega_I &= 0.8297(2).
\end{aligned}
\label{Isiexp}
\end{equation}

\subsection{Finite-size scaling} 
\label{gfobs}

For $N\ge 2$, some relevant observables are obtained from the
correlations of the gauge-invariant bilinear operator $Q_{\bm x}^{ab}$
defined in Eq.~(\ref{qab}). Its two-point correlation function reads
\begin{equation}
G_Q({\bm x},{\bm y}) = \langle {\rm Tr} \,Q_{\bm x} Q_{\bm y} \rangle,
\label{GQdef}
\end{equation}
from which one can also define the corresponding Fourier transform
\begin{equation}
  \widetilde{G}_Q({\bm p})=\frac{1}{L^3}\sum_{{\bm x},{\bm y}}
  e^{i{\bm p}\cdot ({\bm x}-{\bm y})} G_Q({\bm x},{\bm y}), 
\label{GQFourier}
\end{equation}
susceptibility and correlation length:
\begin{eqnarray} 
\chi_q &=& \widetilde{G}_Q({\bm 0}), \label{def-chiq} \\
\xi_q^2 &=& {1\over 4 \sin^2 (\pi/L)}
     {\widetilde{G}_Q({\bm 0})
       - \widetilde{G}_Q({\bm p}_m)\over \widetilde{G}_Q({\bm p}_m)},
\end{eqnarray}
where ${\bm p}_m = (2\pi/L,0,0)$.  In the FSS limit, varying $J$
around the critical point $J_c$ at fixed $K$, $\chi_q$ and $\xi_q$
scale as
\begin{eqnarray}
\chi_q(J,L) &\approx& L^{3-2Y_Q} \, {\cal C}_q(W), 
\label{chiqsca} \\
\xi_q(J,L) &\approx& L\,  {\cal R}_q(W), 
\end{eqnarray}
where $Y_Q$ is the RG
dimension of the spin-two operator $Q_{\bm x}$ and 
\begin{equation}
W = (J-J_c) L^{1/\nu}.
\end{equation}
Scaling corrections decay as $L^{-\omega}$, where $\omega$ is the
leading correction-to-scaling exponent. The ratio
\begin{equation}
  R_q = {\xi_q\over L} \approx {\cal R}_q(W)
  \label{RQdef}
\end{equation}
is RG invariant and can be used to determine $J_c$. Indeed, the data
for different lattice sizes $L$ have a crossing point that coincides
with the critical point for large values of $L$.

In the case of the ${\mathbb Z}_2$-gauge Higgs model there are no
gauge-invariant operator analogous to $Q_{\bm x}$.  The transition may
be probed by studying the fluctuations of the gauge-invariant energy
density, see, e.g., Refs.~\cite{SSN-21,BPV-22-z2h,BPV-24-decQ2}. For
this type of transitions the analysis of the critical gauge-dependent
vector correlations may provide an alternative method to determine the
critical point, beside confirming the Ising$^*$ nature of the
transition. It may also numerically convenient, since numerical
studies based on the energy cumulants are quite demanding.

We also analyze the gauge-dependent correlation function $G_V({\bm
  x},{\bm y})$ in the stochastic gauge fixing approach. If it develops
a nontrivial critical behavior, then it is expected to scale as
(assuming translation invariance for the sake of simplicity)
\begin{eqnarray}
&&G_V({\bm x}-{\bm y},J,L) =L^{-2Y_V} \left[{\cal G}_V({\bm X},W) +
    O(L^{-\omega})\right], \qquad\label{GVscal}
\end{eqnarray}  
where $\nu$ is the length-scale critical exponent and  $Y_V$ is the RG
dimension of the vector operator, which is related to the exponent $\eta$
by
\begin{equation}
  Y_V = {d-2+\eta\over 2} = {1 + \eta\over 2}.
  \label{YV}
\end{equation}
The function ${\cal G}_V$ is expected to be universal, apart from a
multiplicative factor and a normalization of the scaling variables
$W$. It should only depend on the boundary conditions and lattice
shape.  We also consider the corresponding susceptibility $\chi_v$ and
second-moment correlation length $\xi_v$,
\begin{eqnarray}
  \chi_v \equiv  \widetilde{G}_V({\bm 0}),\quad
  \xi_v^2 \equiv {1\over 4 \sin^2 (\pi/L)}
     {\widetilde{G}_V({\bm 0})
       - \widetilde{G}_V({\bm p}_m)\over \widetilde{G}_V({\bm p}_m)},
\quad\label{chixivdef}
\end{eqnarray}
where $\widetilde{G}_V({\bm p})$ is defined as in
Eq.~(\ref{GQFourier}).  Two RG invariant quantities associated with
$G_V({\bm x}, {\bm y})$ are the ratio
\begin{equation}
 R_v \equiv \xi_v/L,
 \label{defR}
\end{equation}
and the Binder parameter
\begin{equation}
  U_v = \frac{\langle[m_{2}^2]\rangle}{\langle[m_{2}]\rangle^2},
  \qquad m_{2} = \frac{1}{L^3} \sum_{{\bm x},{\bm y}} 
  \hat{\bm s}_{\bm x}\cdot \hat{\bm s}_{\bm y}.
\label{uvdef}
\end{equation}
If the correlation function $G_V$ is critical, then we expect 
$R_v$ and $U_v$ to behave as $R_q$, in the FSS limit. For example,
\begin{equation}
  R_v(J,L) = {\cal R}_v(W) + O(L^{-\omega}).
  \label{rvscal}
\end{equation}
Actually, to avoid the
nonuniversal normalization of the argument $W$, one may rewrite 
the Binder parameter in terms of $R_v$:
\begin{equation}
  U_v(J,L) = \widehat{\cal U}(R_v) + O(L^{-\omega}).
  \label{uvsr}
\end{equation}    
Finally, the vector susceptibility is expected to scale as
\begin{equation}
  \chi_v(J,L) \approx L^{2-\eta} \, \widehat{\cal C}_v(R_v).
  \label{chivscal}
\end{equation}

We also consider observables involving the variables $w_{\bm x}$ only.
They allow us to check that the value of $\gamma$ we consider is
sufficiently large so that the quenched system is in the ferromagnetic
phase. To define appropriate observables, it is important to note that
the gauge-fixed theory is gauge invariant under the transformations
(\ref{gaugetr-2}).  Therefore, we consider the so-called replica
observables, as commonly done in the analysis of random quenched
systems.  If $w^{(1)}_{\bm x}$ and $w^{(2)}_{\bm x}$ are two different
system variables distributed with probability $e^{-H_w}/Z_w$ with the
same disorder distribution (same $\sigma_{{\bm x},\mu}$, ${\bm s}_{\bm
  x}$ in the present context), we define the overlap variable as the
product
\begin{equation}
  O_{\bm x} = w^{(1)}_{\bm x} w^{(2)}_{\bm x}.
  \label{overlap}
\end{equation} 
The corresponding susceptibility is
\begin{equation}
  \chi_o = L^{-3}
  \langle \bigl[\; \bigl(\sum_{\bm x} O_{\bm x}\bigr)^2\;\bigr]\rangle, 
  \label{chiudef}
\end{equation}
while the  Binder parameter is defined as 
\begin{equation}
  U_o = 
\frac{\langle[n_{2}^2]\rangle}{\langle[ n_{2}]\rangle^2},
  \qquad n_{2} = L^{-3} \sum_{{\bm x},{\bm y}} 
  O_{\bm x} \, O_{\bm y}.
  \label{uoverlap}
\end{equation}
In the ferromagnetic phase [$\gamma > \gamma_c(K,J)$] we expect the
finite-size behavior
\begin{equation}
\chi_o\sim L^3,\qquad U_o=1 + O(L^{-3}).
\label{ferrbeh}
\end{equation}

\subsection{Results for ${\mathbb Z}_2$-gauge $N=3$
  vector model}
\label{n3res}

\begin{figure}[tb]
\centering
\includegraphics[width=0.85\columnwidth, clip]{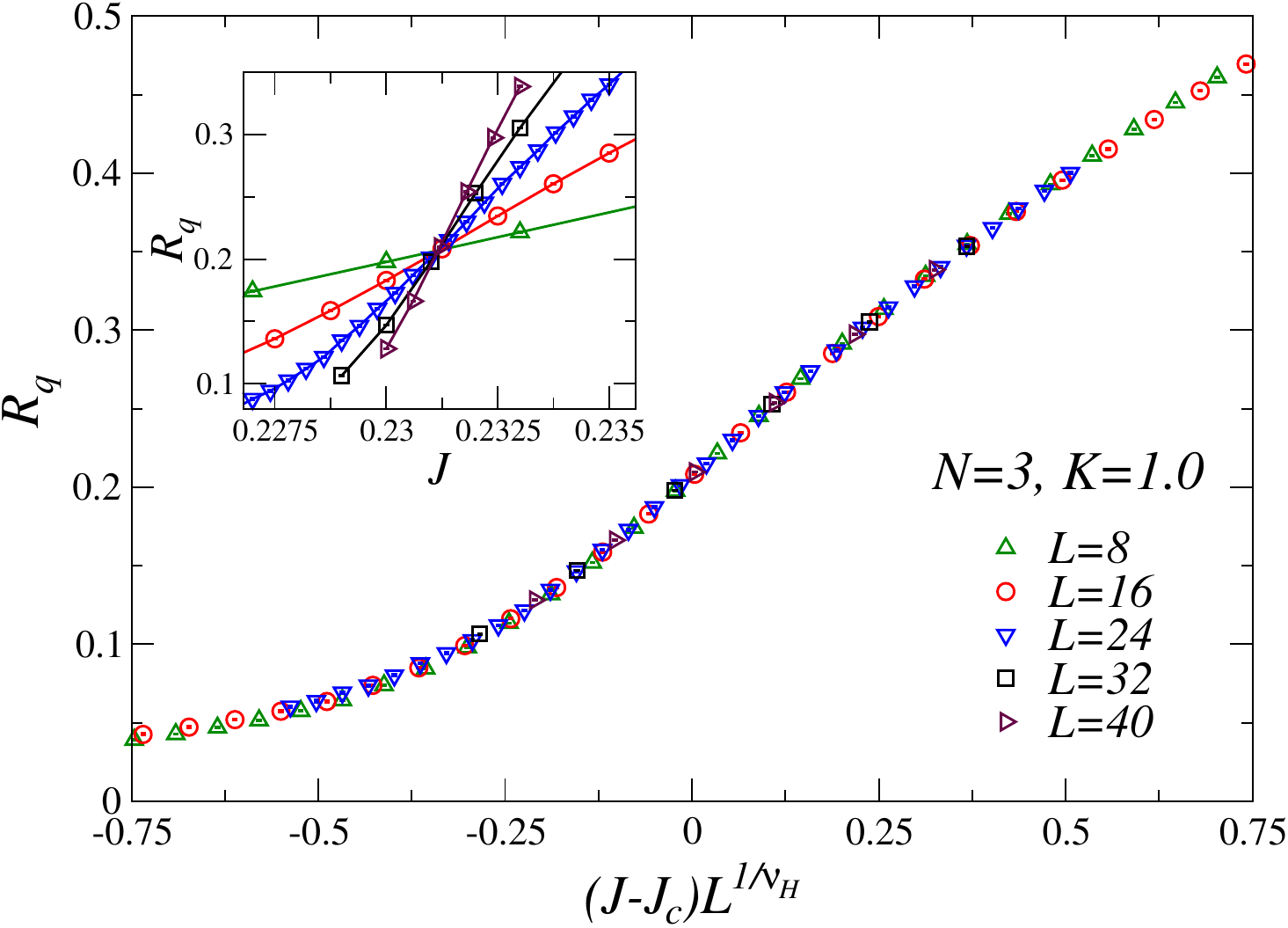}
\caption{Scaling plot of the ratio $R_q$ for the ${\mathbb Z}_2$-gauge
  $N=3$ vector model at $K=1$, with periodic boundary conditions. We
  plot $R_q$ versus $W=(J-J_c)L^{1/\nu}$ with $J_c=0.23118(3)$ and
  $\nu = \nu_H = 0.71164$. In the inset we report $R_q$ as a function
  of $J$.}
\label{RQN3}
\end{figure}

\begin{figure}[tb]
\centering
\includegraphics[width=0.85\columnwidth, clip]{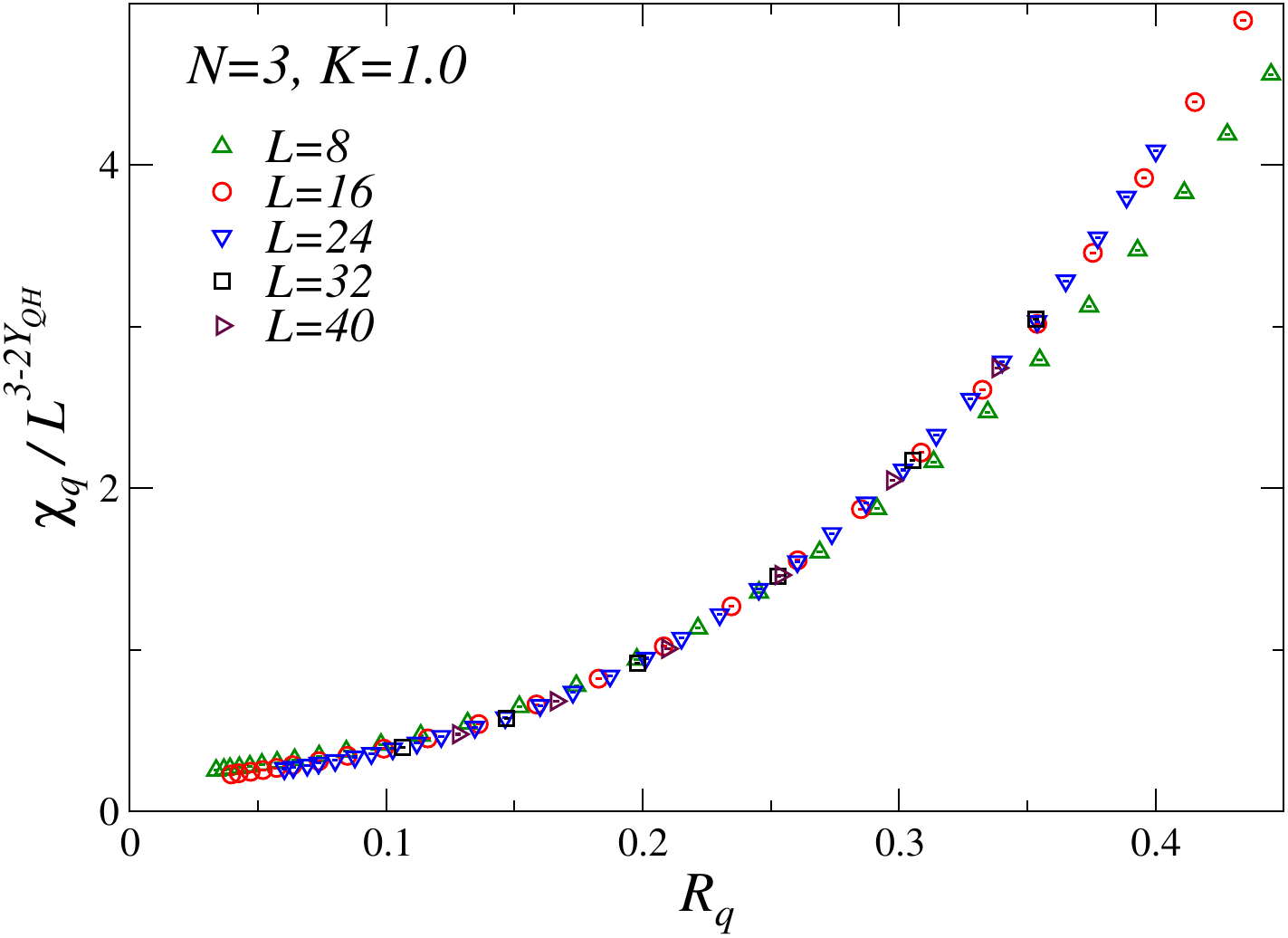}
\caption{Scaling plot of $\chi_q$ for the ${\mathbb Z}_2$-gauge $N=3$
  vector model at $K=1$, with periodic boundary conditions.  We report
  $L^{-3+2Y_{Q}} \chi_q$ versus $R_q$, using the RG dimension $Y_{Q}$
  for the Heisenberg universality class: $Y_Q = Y_{QH} = 1.2094$.}
\label{chiQN3}
\end{figure}

We now present our numerical FSS analyses for the ${\mathbb
  Z}_2$-gauge model with $N=3$. We consider periodic boundary
conditions and perform simulations along the line $K=1$. For $J\approx
0.23$ the model undergoes a continuous O(3)$^*$ transition.  This is
clearly demonstrated by the FSS behavior of ratio $R_q$ and of the
susceptibility $\chi_q$ defined in Eqs.~(\ref{RQdef}) and
(\ref{def-chiq}).  In Fig.~\ref{RQN3} we show the results for
$R_q$. Data have a clear crossing point and show an excellent scaling
if we set $\nu = \nu_H$, confirming the Heisenberg nature of the
transition.  Fits of the data fixing $\nu=\nu_H$ and $\omega=\omega_H$
provide an accurate estimate of the critical point
$J_c=0.23118(3)$. Note that $J_c$ is quite close to the critical value
in the Heisenberg model ($K\to\infty$)
$J_H=0.2310010(7)$~\cite{DBN-05, BFMM-96}, indicating that the
transition line is almost parallel to the $K$ axis.  The scaling
behavior of $\chi_q$, see Fig.~\ref{chiQN3}, is fully consistent with
Eq.~(\ref{chiqsca}), using the RG dimension $Y_Q = Y_{QH}$ of the
spin-2 operator at the O(3) vector fixed point, again in agreement
with the general scenario.

We now analyze the vector correlation $G_V$, defined in
Eq.~(\ref{gvgf}). As discussed in Sec.~\ref{gaufix}, the parameter
$\gamma$ of the ancillary Hamiltonian $H_w$, must be chosen so that
the ancillary system is in the ferromagnetic phase. This can be easily
checked by looking at the behavior of the overlap observables defined
in Eqs.~(\ref{chiudef}) and (\ref{uoverlap}). For $\gamma=0.3$, with
increasing $L$, we find that $\chi_o/L^3\approx 0.6755$ and that the
difference $U_o-1$ vanishes as $L^{-3}$, as expected for a
ferromagnetic phase. Therefore, in the following we compute the vector
observables fixing $\gamma = 0.3$.

\begin{figure}[tb]
\centering
\includegraphics[width=0.85\columnwidth, clip]{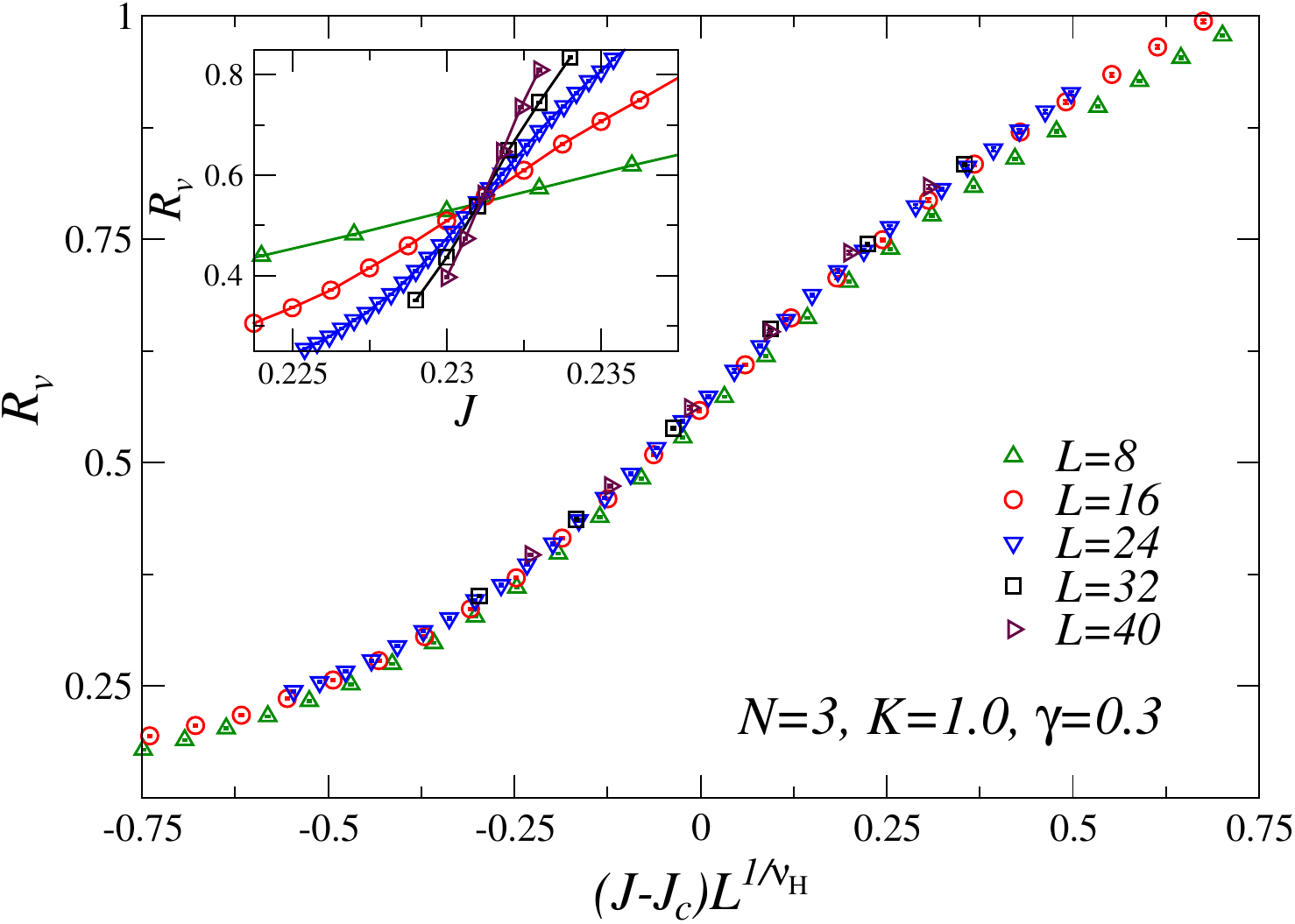}
\caption{Ratio $R_v=\xi_v/L$ for the ${\mathbb Z}_2$-gauge $N=3$
  vector model at $K=1$, using the stochastic gauge fixing with
  $\gamma=0.3$. Plot of $R_v$ as a function of $W=(J-J_c)L^{1/\nu_H}$
  and as a function of $J$ (see inset). We fix $\nu = \nu_H =
  0.71164$. }
\label{RVN3}
\end{figure}

\begin{figure}[tb]
\centering
\includegraphics[width=0.9\columnwidth, clip]{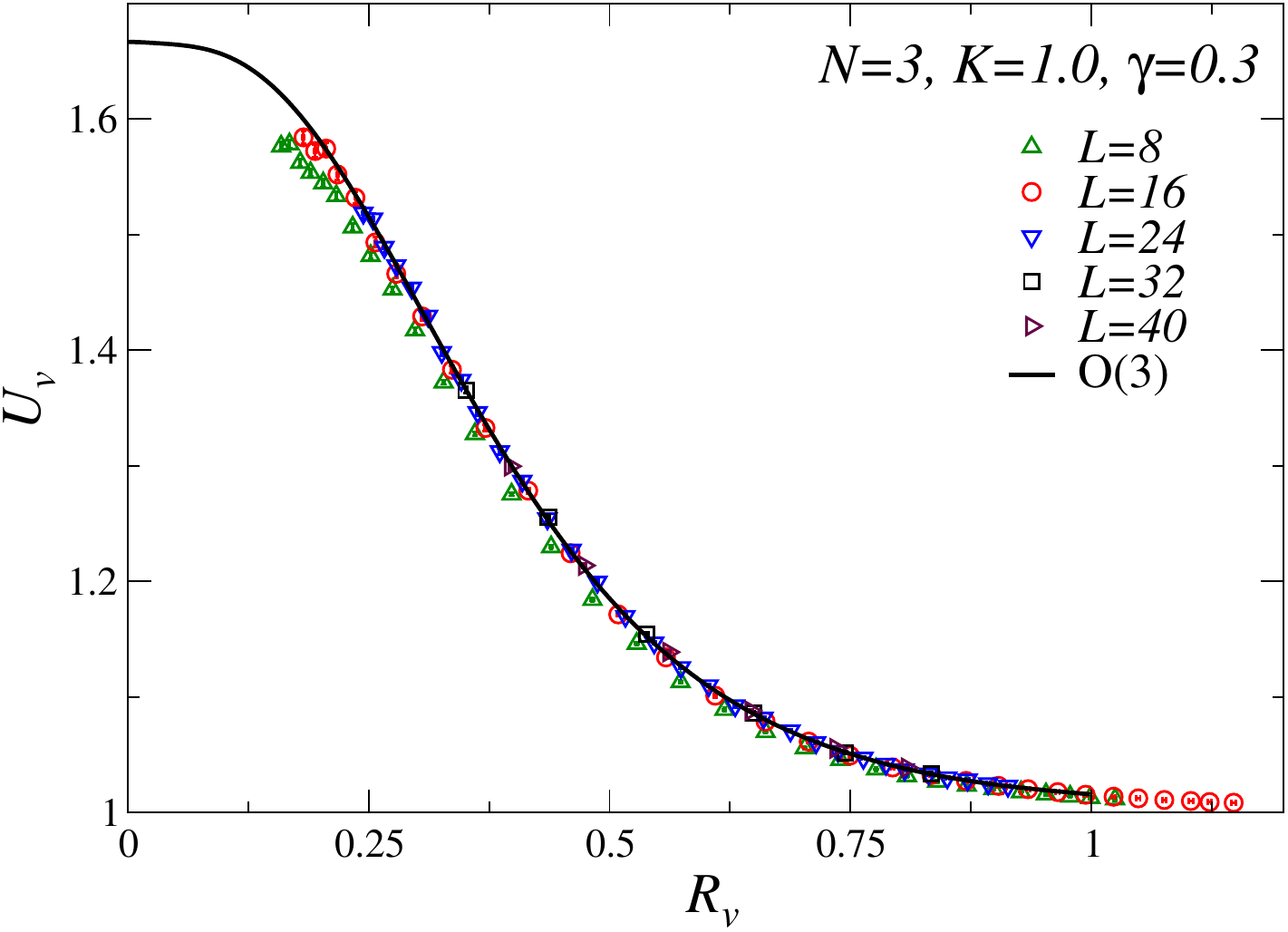}
\caption{$U_v$ as a function of $R_v$ for the ${\mathbb Z}_2$-gauge
  $N=3$ vector model at $K=1$, using the stochastic gauge fixing with
  $\gamma=0.3$, for different values of the lattice size $L$ and of
  the coupling $J$.  The data follow the FSS behavior
  (\ref{uvsr}). Moreover, the asymptotic FSS curve is consistent with
  the universal FSS curve for the standard O(3) vector model with
  periodic boundary conditions (solid line). For the O(3) curve we use
  the parametrization reported in Ref.~\cite{BPV-21-ccb}.  }
\label{UVRVN3}
\end{figure}

\begin{figure}[tb]
\centering \includegraphics[width=0.85\columnwidth,
  clip]{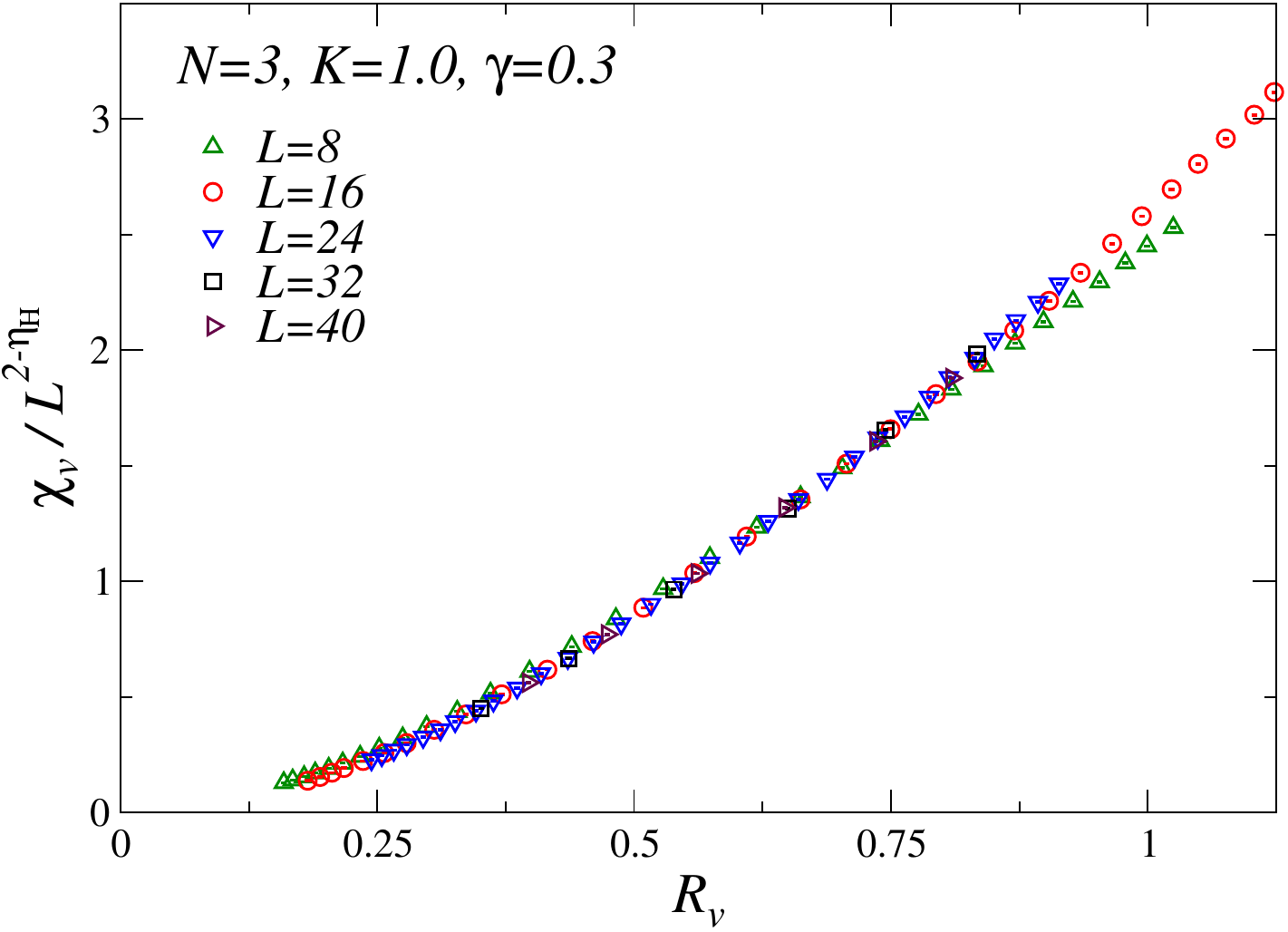}
\caption{Vector susceptibility $\chi_v$ for the ${\mathbb Z}_2$-gauge
  $N=3$ vector model at $K=1$, using the stochastic gauge fixing with
  $\gamma=0.3$. The excellent scaling of the ratio
  $\chi_v/L^{2-\eta_H}$ with $\eta_H = 0.03784$ confirms that along
  the O(3)$^*$ transition line the correlation function $G_V$ behaves
  as the vector correlation function in the O(3) vector model.  }
\label{chiVN3}
\end{figure}

Results for $R_v=\xi_v/L$ are reported in Fig.~\ref{RVN3}. The
correlation length increases as $L$, as expected for a critical
observable and shows a crossing point at a value of $J$ that is very
close to the critical point $J_c$ obtained fitting the gauge-invariant
observable $R_q$. More precisely, fits of $R_v$ give $J_c=0.23128(8)$,
which is in good agreement with the estimate $J_c = 0.23118(3)$
determined above. The $R_v$ data also show an excellent FSS behavior
if we use the Heisenberg estimate of the correlation length exponent,
see Fig.~\ref{RVN3}.  To obtain a robust check that $G_V$ has the same
critical behavior as the Heisenberg correlation function, in
Fig.~\ref{UVRVN3} we report $U_v$ versus $R_v$ together with the
universal scaling function for the analogous vector quantities
computed in the Heisenberg model (the curve is taken from
Ref.~\cite{BPV-21-ccb}).  Also in this case the agreement is
excellent. Finally, we determine the scaling behavior of $\chi_v$. The
vector susceptibility scales very nicely according to
Eq.~(\ref{chivscal}), if we set $\eta = \eta_H$, where $\eta_H$ is the
Heisenberg exponent for vector correlations.

In conclusion, the above FSS analyses demonstrate the effectiveness of
the stochastic gauge fixing outlined in Sec.~\ref{gaufix}. Indeed, it
allows us to identify a critical gauge-dependent vector field that
orders at the transition, and therefore can be taken as the order
parameter for the spontaneous breaking of the O($N$) symmetry to
O($N-1$). In turn, this allows us to obtain a complete mapping of the
RG operators of the O($N)$ and O($N)^*$ transitions.

\subsection{Results for the ${\mathbb Z}_2$-gauge Higgs model}
\label{n1res}

\begin{figure}[tb]
\centering
\includegraphics[width=0.85\columnwidth, clip]{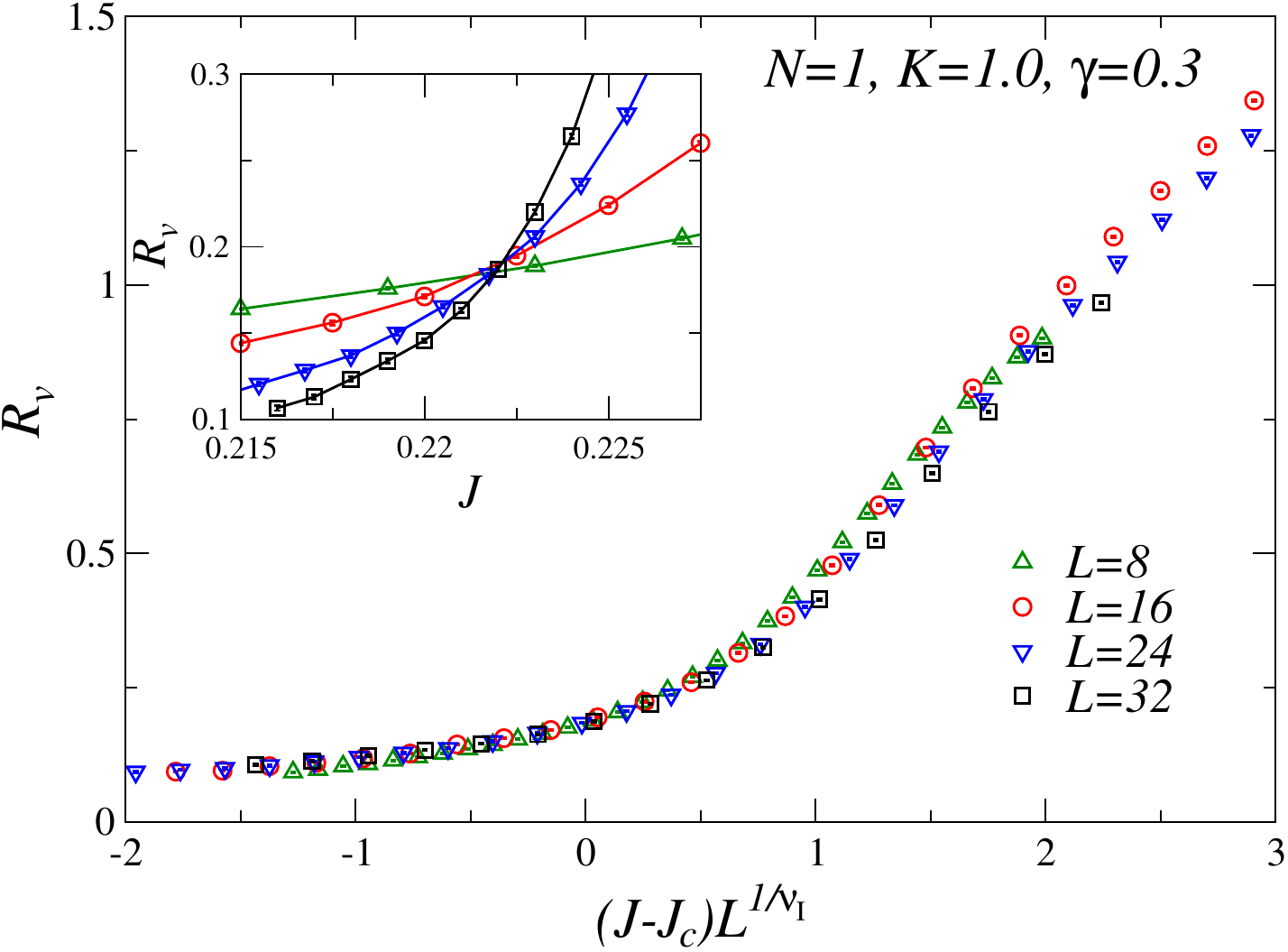}
\caption{Ratio $R_v=\xi_v/L$ for the ${\mathbb Z}_2$-gauge Higgs model
  at $K=1$, using the stochastic gauge fixing with $\gamma=0.3$.  We
  plot $R_v$ versus $W=(J-J_c)L^{1/\nu_I}$ using $\nu = \nu_I =
  0.629971$ and the estimate $J_c=0.22185(10)$, obtained by biased
  fits of $R_v$ using $\nu=\nu_I$.  The inset shows the same data
  versus $J$.  }
\label{RVN1}
\end{figure}

\begin{figure}[tb]
\centering \includegraphics[width=0.85\columnwidth,
  clip]{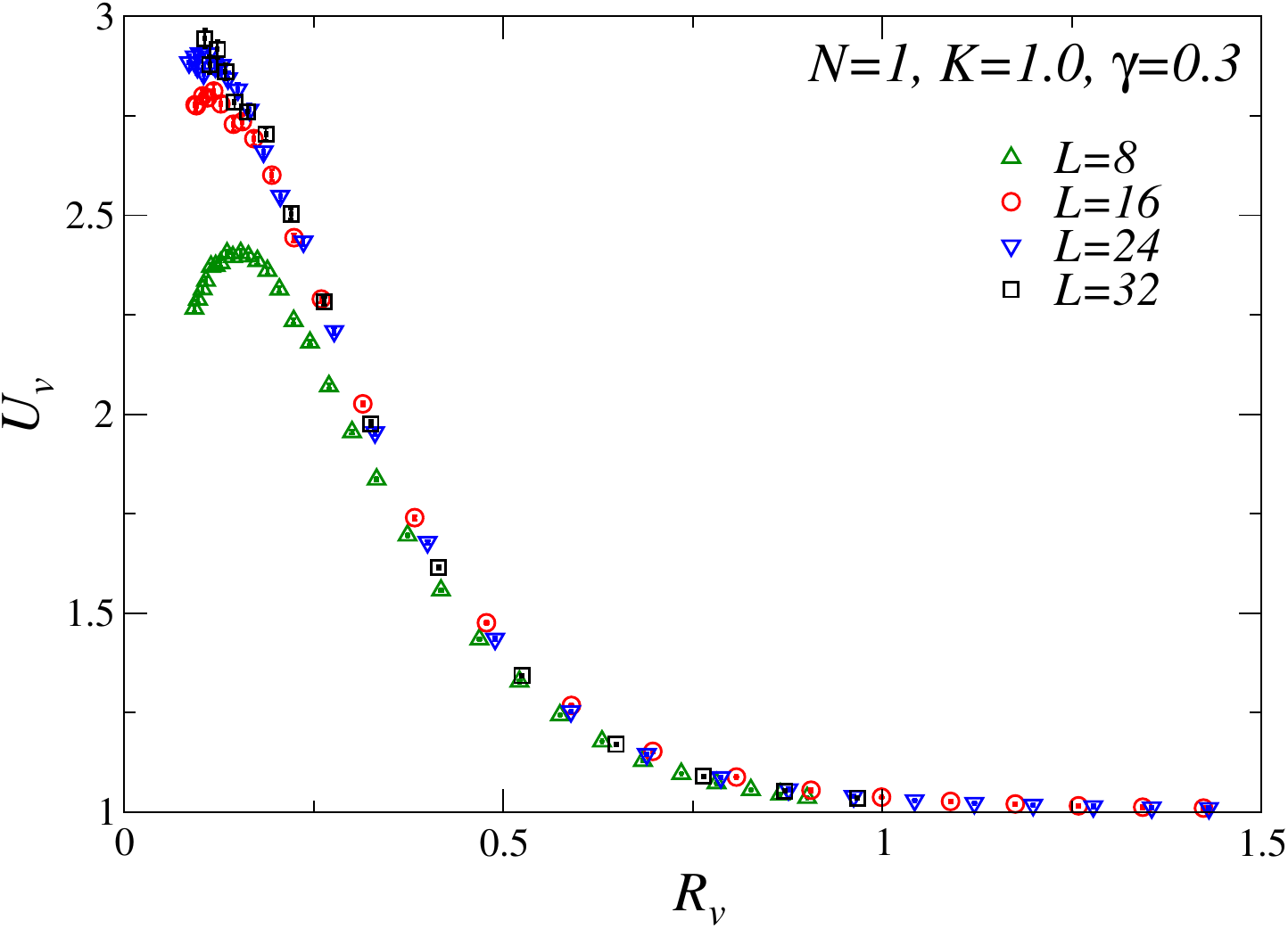}
\caption{Vector Binder parameter $U_v$ as a function of $R_v$ for the
  ${\mathbb Z}_2$-gauge Higgs model at $K=1$, using the stochastic
  gauge fixing with $\gamma=0.3$. }
\label{UVRVN1}
\end{figure}

\begin{figure}[tb]
\centering
\includegraphics[width=0.9\columnwidth, clip]{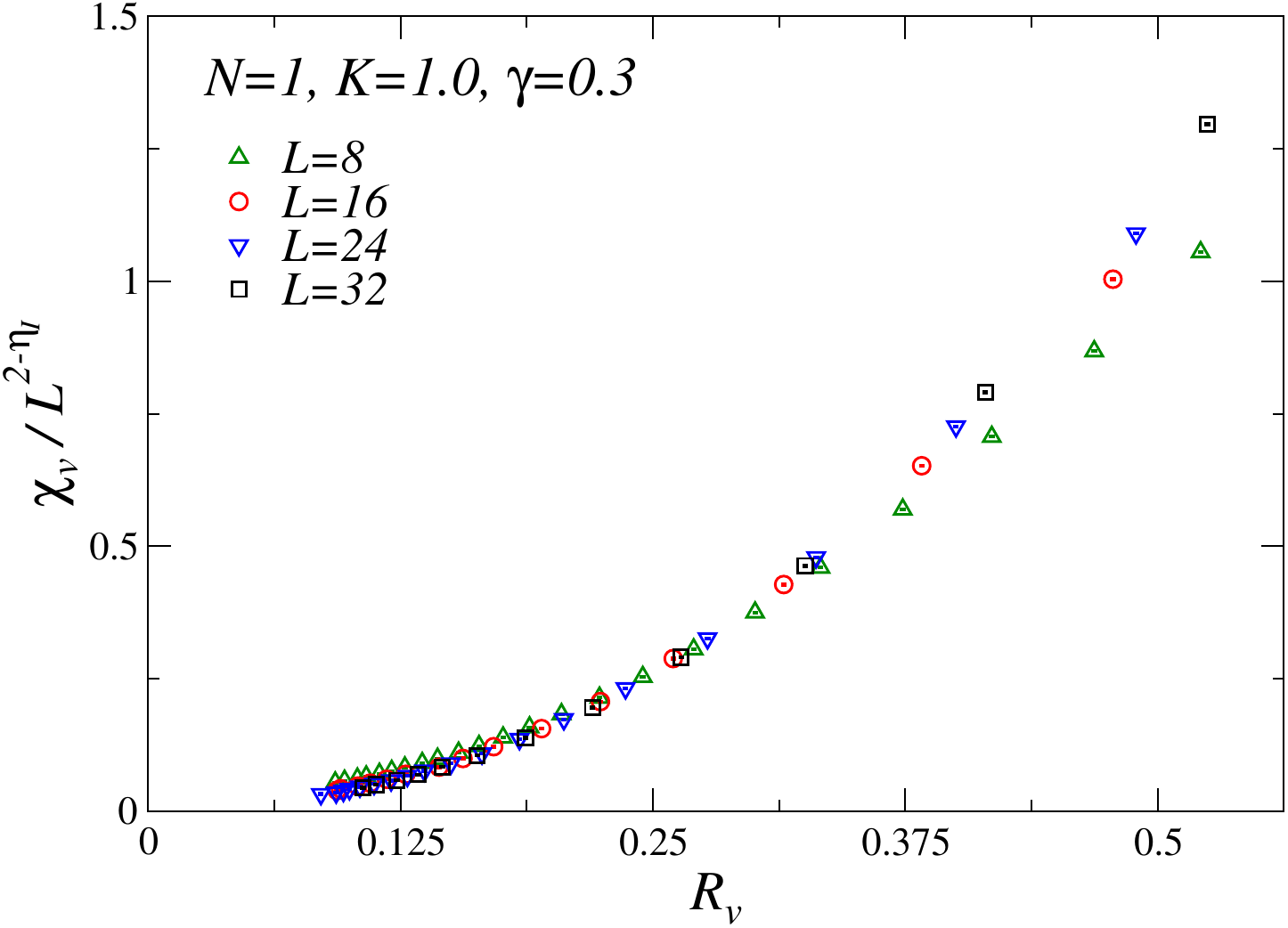}
\caption{Vector susceptibility $\chi_v$ for the ${\mathbb Z}_2$-gauge
  Higgs model at $K=1$, using the stochastic gauge fixing with
  $\gamma=0.3$. We set $\eta = \eta_I = 0.036298$.  The excellent
  scaling confirms that $G_V$ along the Ising$^*$ transition line
  behaves as the standard Ising spin-spin correlation function. }
\label{chiVN1}
\end{figure}

We now present results for the ${\mathbb Z}_2$-gauge Higgs model.  We
study the critical behavior at the Ising$^*$ transition along the line
$K=1$, using the stochastic gauge fixing, to uncover the
gauge-dependent spin critical fluctuations.  We use open boundary
conditions to avoid the long autocorrelation times associated with the
Polyakov lines.\footnote{Note that definitions reported in
Sec.~\ref{gfobs} are valid both for open and periodic boundary
conditions.}  We fix $\gamma=0.3$: $\chi_o$ and $U_o$ close the
transition behave as in Eq.~(\ref{ferrbeh}), confirming that the
ancillary system is ferromagnetic for this value of $\gamma$.

The correlation $G_V$ behaves as the spin correlation function at the
standard Ising transition. This is confirmed by the data of
$R_v=\xi_v/L$ shown in Fig.~\ref{RVN1}. The correlation length $\xi_v$
shows an excellent scaling if one uses the Ising critical exponent
$\nu_I$.  A fit of the data provides the accurate estimate
$J_c=0.22185(10)$, which is again quite close to the critical value in
the $K\to\infty$ limit, i.e., $J_{Is}=0.221654626(5)$ of the standard
Ising model~\cite{FXL-18}. The critical nature of the gauge-dependent
correlations is further confirmed by the plot of $U_v$ versus $R_v$
shown in Fig.~\ref{UVRVN1}.  Data approach an asymptotic scaling
curve, as predicted by the FSS Eq.~(\ref{uvsr}). Finally, the scaling
behavior of the susceptibility $\chi_v$ is definitely consistent with
Eq.~(\ref{chivscal}), setting $\eta = \eta_I$, where $\eta_I$ is the
Ising exponent.  Data reported in Fig.~\ref{chiVN1} show an excellent
scaling.

The above FSS results show again the effectiveness of the stochastic
gauge fixing outlined in Sec.~\ref{gaufix}, which allows us to
identify the universal gauge-dependent spin correlations which
characterize standard Ising transitions, see, e.g., Ref.~\cite{PV-02}.

\subsection{Transitions of the ancillary quenched model}
\label{ancres}

\begin{figure}[tb]
\centering
\includegraphics[width=0.9\columnwidth, clip]{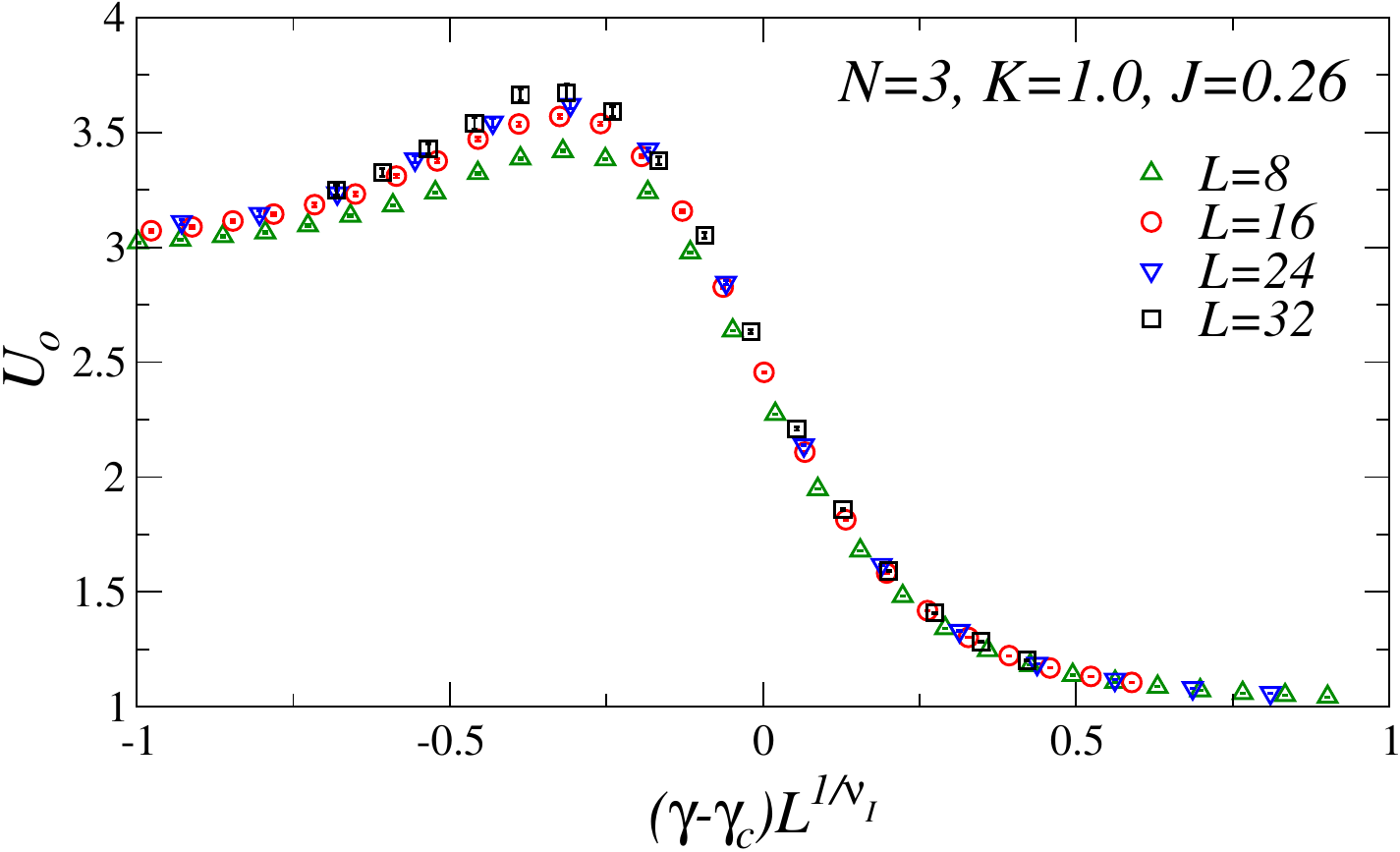}
\includegraphics[width=0.9\columnwidth, clip]{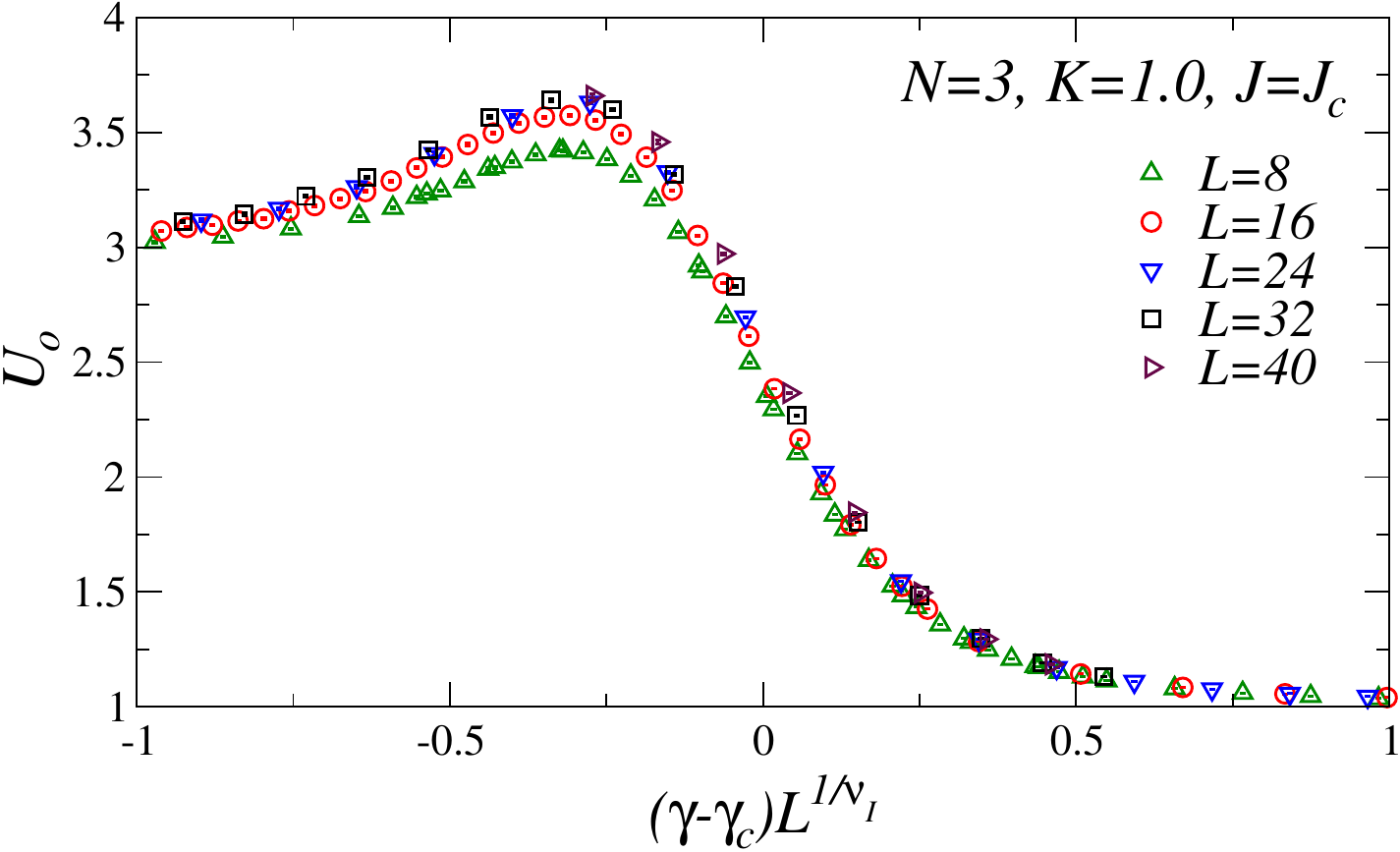}
\includegraphics[width=0.9\columnwidth, clip]{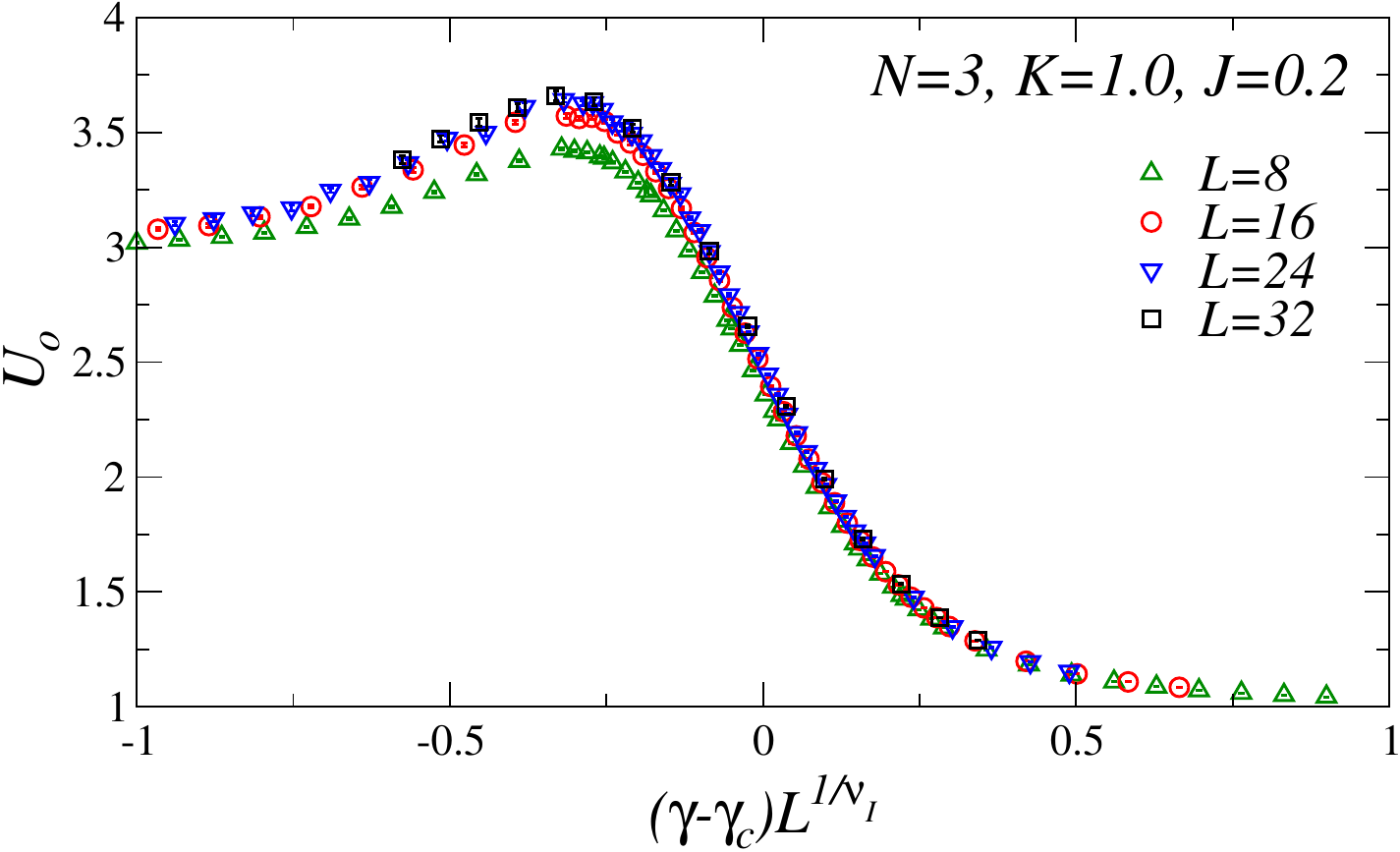}
\caption{Overlap Binder parameter $U_o$ for $N=3$, $K=1$, and three
  different values of $J$ as a function of $(\gamma - \gamma_c)
  L^{1/\nu_I}$, where $\nu_I = 0.629971$ is the Ising critical
  exponent. Results for: (top) $J = 0.26 > J_c$ (O phase) with
  $\gamma_c=0.22178$; (middle) $J = 0.23118\approx J_c$ with
  $\gamma_c=0.22178$; (bottom) $J=0.2<J_c$ (DO phase) with
  $\gamma_c=0.22185$.  }
\label{UoJ}
\end{figure}

In the previous subsections we focused on the critical behavior of the
vector correlations defined by using the stochastic gauge fixing, at
the transition point of the original gauge-invariant lattice model.
We now focus on the ancillary quenched random-bond model.  For any $J$
and $K$ it undergoes a quenched transition at $\gamma_c(J,K)$, which
separates a small-$\gamma$ disordered phase from a large-$\gamma$
ordered phase. Here we wish to investigate the nature of this
transition in the different phases of the gauge-invariant lattice
model.  We only report results for $N=3$, but the general picture
should be valid for any $N$, including the gauge-Higgs model with
$N=1$. We focus on the behavior along the line $K = 1$, considering
three values of $J$: (i) $J = 0.2 < J_c$ in the DO phase; (ii)
$J=J_c\approx 0.2312$, where the gauge-invariant model is critical;
(iii) $J = 0.26 >J_c$, in the O phase.

For $J = 0.2$, in the DO phase, the overlap variables show the
presence of a transition for $\gamma \approx 0.222$, with critical
exponents that are definitely consistent with the Ising values.  This
is clearly confirmed by the data shown in the lower panel of
Fig.~\ref{UoJ}, where we plot the overlap Binder parameter as a
function of $(\gamma-\gamma_c)L^{1/\nu_I}$, with Ising exponent
$\nu_I$ (we use $\gamma_c=0.22185(3)$ as obtained from the crossing
point of the overlap Binder parameter).  Moreover, the data of the
overlap susceptibility $\chi_o$ at $\gamma_c$ (not shown) are
consistent with the behavior $\chi_o\sim L^{2-\eta_{qI}}$, where
$\eta_{qI}=1+2\eta_I=1.072596(4)$ is the Ising exponent associated
with the RG dimension of the overlap variable.  We expect these
results to be the same in the whole DO phase. This is confirmed by the
results of Ref.~\cite{CP-19}, that also observed a pure Ising critical
behavior on the line $J=0$ in the DO phase.

These results indicate that the type of bond disorder that occurs in
the DO phase does not destabilize the pure Ising fixed point.  This is
different from what occurs in generic random-bond models with
spatially-uncorrelated bimodal or Gaussian bond distributions.  In
that case, the Harris criterion~\cite{Harris-74} predicts the
instability of the pure Ising fixed point against quenched disorder,
due to the positive value of the 3D Ising specific-heat exponent.
Indeed, in the presence of spatially-uncorrelated quenched disorder,
one expects that the ferromagnetic critical behavior belongs to
another universality class, the so-called randomly-diluted Ising
universality class, with critical exponents \cite{HPPV-07,PV-02}
$\nu_{rI} = 0.683(2)$ and $\eta_{rI}=0.036(1)$.  The observed pure
Ising critical behavior, i.e., the apparent stability of the pure
Ising fixed point, is related to the different nature of the quenched
disorder in the present model, that corresponds to a
topologically-ordered phase.

In the DO phase, for $J<J_c$, we do not expect the stochastic gauge
fixing to lead to any ordering of the vector correlation $G_V$ defined
in Eq.~(\ref{gvgf}), for any $\gamma$, because the condensation of the
gauge-dependent vector field should be accompanied by the condensation
of the gauge-invariant bilinear operator $Q_{\bm x}$, which only
occurs for $J>J_c$ and is not affected by the stochastic gauge
fixing. Numerical data confirm this general picture.

\begin{figure}[tb]
\centering \includegraphics[width=0.85\columnwidth,
  clip]{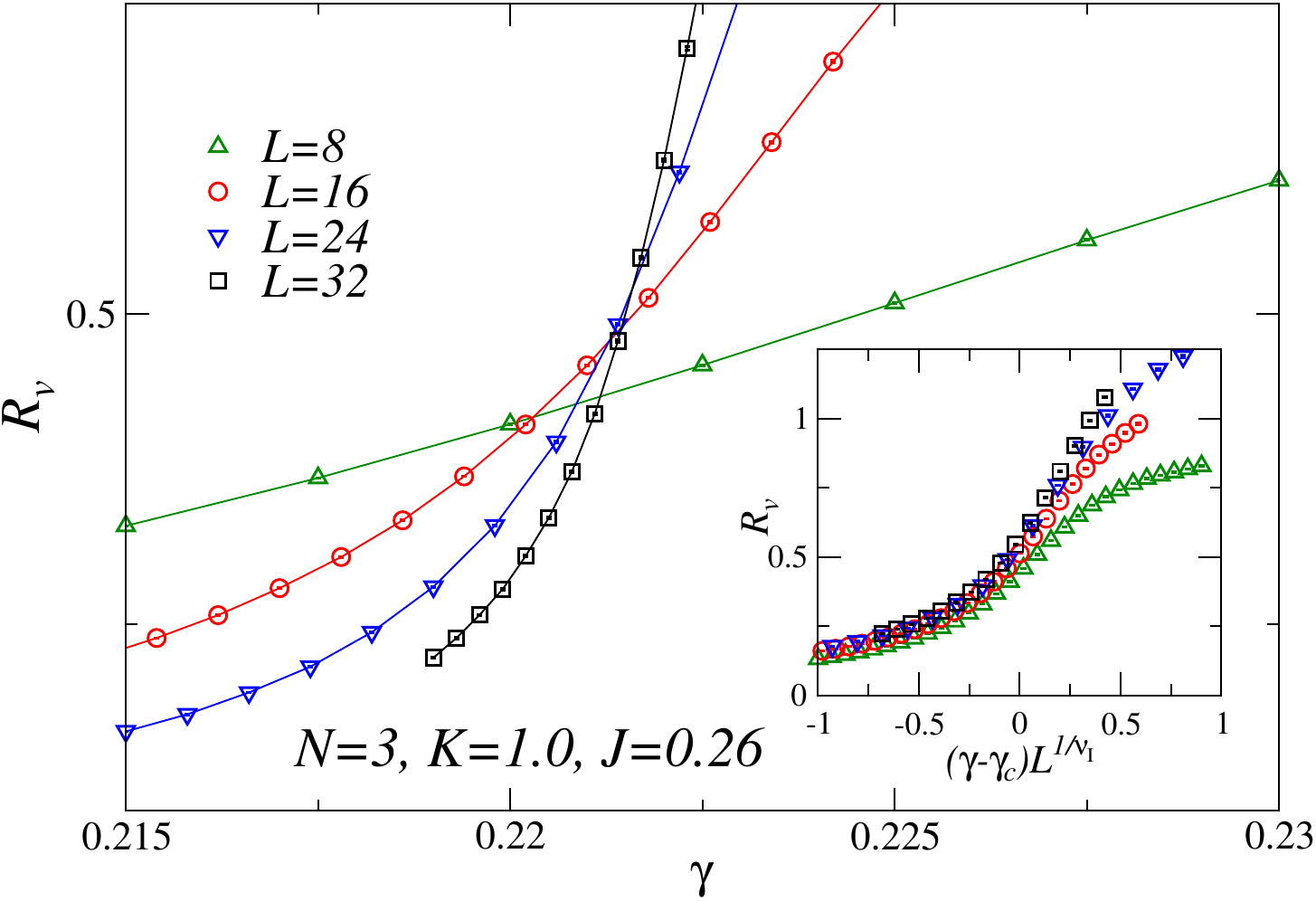}
\caption{The length-scale ratio $R_v=\xi_v/L$ defined in
  Eq.~(\ref{chixivdef}) as a function of the stochastic gauge fixing
  parameter $\gamma$, for $N=3$, $K=1$, and $J=0.26>J_c$. In the inset
  we report a scaling plot using the Ising critical exponent $\nu_I =
  0.629971$.  }
\label{RvJltJc}
\end{figure}

\begin{figure}[tb]
\centering
\includegraphics[width=0.85\columnwidth, clip]{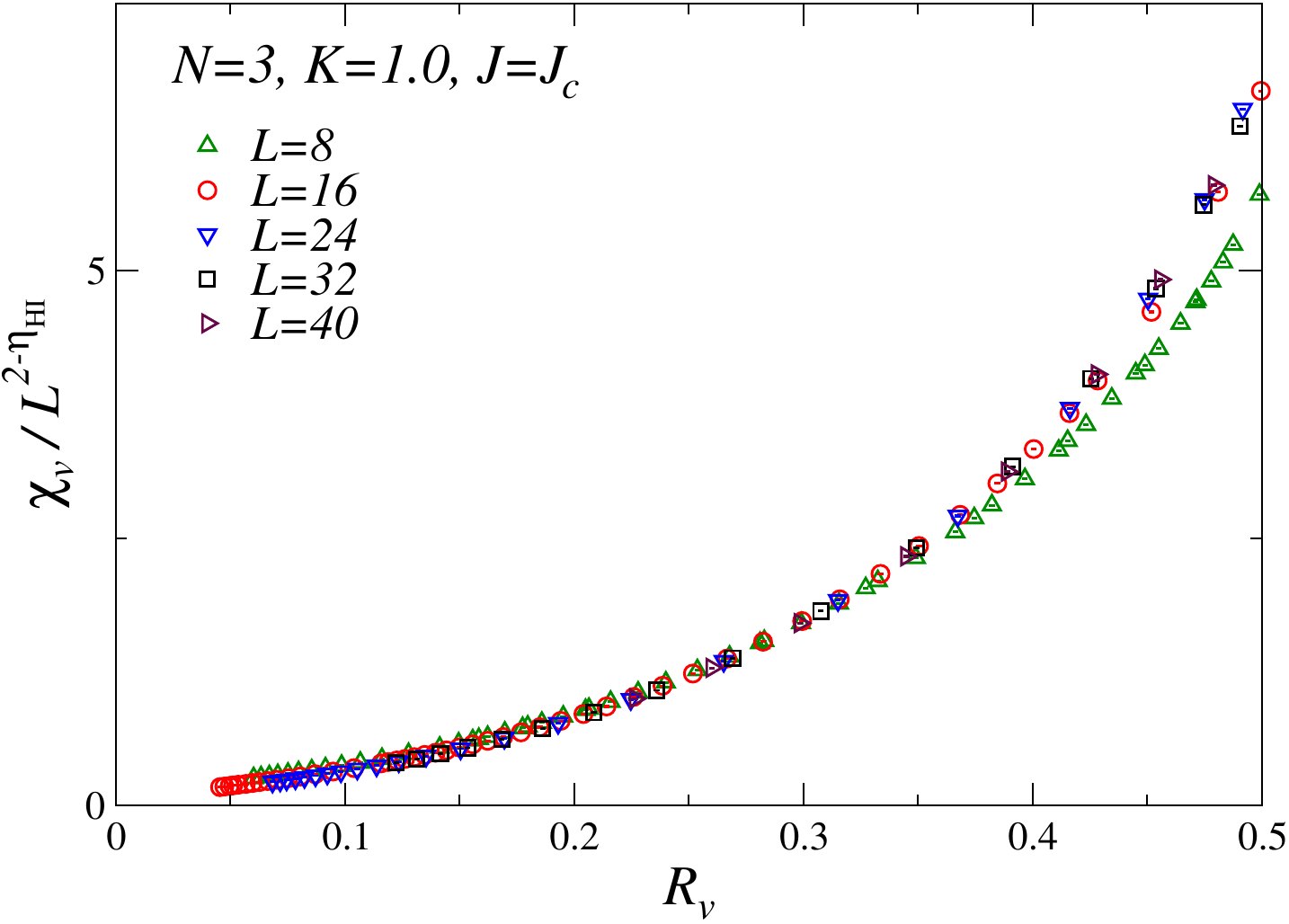}
\includegraphics[width=0.85\columnwidth, clip]{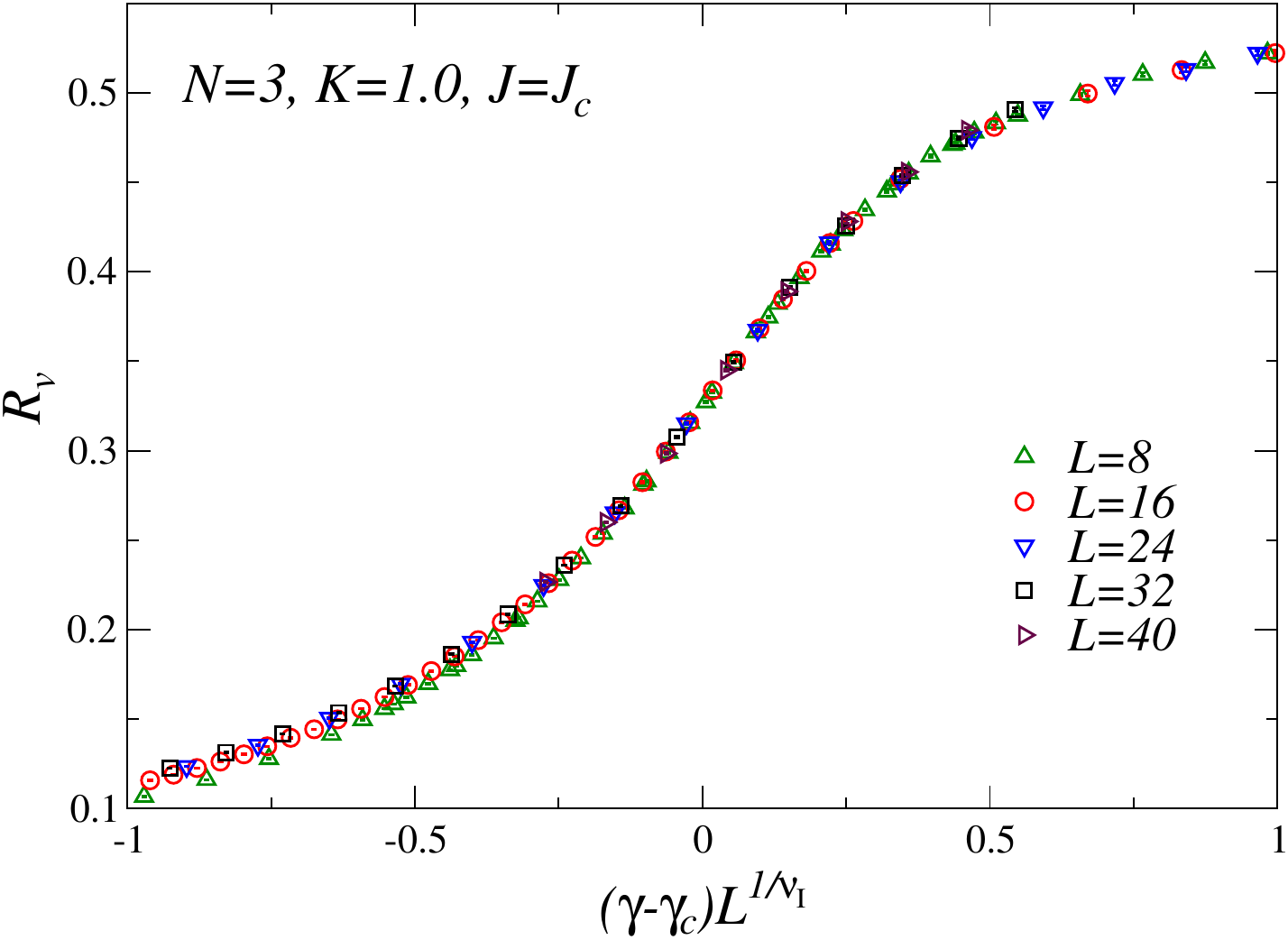}
\caption{Scaling of the $R_v=\xi_v/L$ and of the susceptibility
  $\chi_v$, see Eq.~(\ref{chixivdef}), as a function of the stochastic
  gauge fixing parameter $\gamma$.  Results for $N=3$, $K=1$, and
  $J=0.2312\approx J_c$. We plot $R_v$ versus
  $(\gamma-\gamma_c)L^{1/\nu_I}$ with $\gamma_c=0.22178$ (bottom), and
  $\chi_v/L^{2-\eta_{HI}}$ versus $R_v$ with $\eta_{HI}=1+\eta_I +
  \eta_H =1.07414$ (top).}
\label{chiRvJ}
\end{figure}

Let us now turn our attention to the O phase.  For $J\to \infty$, the
arguments reported at the end of Sec.~\ref{sec3.2} indicate that the
$w_{\bm x}$ variables behave as Ising variables. Therefore, for
$\gamma = \gamma_c$ overlap variables should behave as in the Ising
case. Results for $J = 0.26 > J_c$ are fully consistent with this
picture, as demonstrated by the data reported in the top panel of
Fig.~\ref{UoJ}. The overlap Binder parameter shows the expected FSS if
we set $\nu = \nu_I$, where $\nu_I$ is the Ising critical exponent. In
the plot we use $\gamma_c=0.22178(6)$, as obtained by biased fits of
the data setting $\nu = \nu_I$.  For $J\to \infty$, $G_V$ behaves as
the spin-spin correlation function in the Ising model, see
Sec.~\ref{sec3.2}.  It is natural to expect the same behavior in the
whole ordered phase.  The critical nature of $G_V$ for $\gamma =
\gamma_c$ in the O phase is demonstrated by the data for $R_v$
computed for $J=0.26$ reported in Fig.~\ref{RvJltJc}.  There is a
clear crossing point, indicating that $\xi_v\sim L$ for $\gamma =
\gamma_c$, at approximately the same value of $\gamma_c$ obtained from
the analysis of the overlap variables. However, in the ordered phase,
scaling corrections are large. They are probably related to the nearby
presence of the critical point $J=J_c\approx 0.23$, where also the
disorder is critical. We have also investigated the behavior of the
susceptibility $\chi_v$, which is expected to scale $\chi_v \sim
L^{2-\eta_I} \sim L^{1.96}$. Data show significant scaling
corrections. Fits of $\chi_v$ versus $L^\kappa$ at the critical point
$\gamma = \gamma_c$ provide estimates of $\kappa$ that increase as
lower-$L$ data are discarded. If we consider only data with $L\ge 16$,
this behavior is consistent with $\eta=\eta_I$ and the presence of
large scaling corrections parametrized by the exponent
$\omega=\omega_I$. These large scaling corrections are probably due to
the spin modes being not fully magnetized, given the relatively small
lattice sizes we consider and the small distance between $J=0.26$ and
the critical point.

Finally, let us consider the behavior for $J=J_c$. In this case we are
dealing with a multicritical point, where both the gauge-invariant
observables and the ancillary variables $w_{\bm x}$ are critical.  As
we already discussed, both in the DO and O phases, the variables
$w_{\bm x}$ show Ising criticality. Apparently, the ordering of the
spin degrees of freedom has little influence on the behavior of the
gauge variables $w_{\bm x}$. This is also supported by the estimates
of the critical point: $\gamma_c = 0.22185(3)$ for $J=0.2$,
$\gamma_c=0.22178(3)$ at $J=J_c$, and $\gamma_c=0.22178(6)$ at
$J=0.26$. We thus expect a pure Ising behavior of the overlap
correlations along the whole line $K=1$, from $J=0$~\cite{CP-19} to
$J=\infty$, including the critical point $J=J_c$. More generally,
Ising behavior should occur in the DO and O phases, including the DO-O
transition line.  This is in agreement with the exact results for
$K=\infty$ reported at the end of Sec.~\ref{sec3.2}. Indeed, in this
limit the overlap variables behave as Ising variables for any value of
$J$, because of the factorization of the partition function.
Numerical results are fully consistent with this picture, as
demonstrated by the data reported in the middle panel of
Fig.~\ref{UoJ}.

Let us finally discuss the behavior of $G_V$ at the critical point. In
the limit $K \to \infty$, the correlation function factorizes as
indicated in Eq.~(\ref{GV-factorized}). In infinite volume at fixed
$J=J_c$ we can rewrite it as
\begin{equation} 
G_V({\bm x}) = {Z\over r^{1 + \eta_H}} G_{\rm Is}({\bm x}),
\end{equation} 
where $G_{\rm Is}({\bm x})$ is the Ising correlation function and
$r=|{\bm x}|$.  We thus predict that $\xi_v$ behaves as an Ising
correlation length, while the susceptibility $\chi_v$ is expected to
scale as $L^{2 - \eta_{HI}}$ with $\eta_{HI} = 1 + \eta_H + \eta_I =
1.07414(10)$.  We expect these results to hold on the whole DO-O
transition line.  This is is fully supported by the numerical
data. For instance, see the lower panel of Fig.~\ref{chiRvJ}, we
observe an excellent scaling if we plot $R_v$ in terms of $W =
(\gamma-\gamma_c) L^{1/\nu_I}$, using the Ising critical
exponent. Analogously, the susceptibility $\chi_v$ scale as $L^{2 -
  \eta_{HI}}$ with $\eta_{HI} = 1.07414(10)$, see the upper panel of
Fig.~\ref{chiRvJ}.

It is interesting to observe that, since the point $J=J_c$, $\gamma=
\gamma_c(J_c)$ is a multicritical, it is characterized by two
different length scales with exponents $\nu_I$ and $\nu_H$,
respectively. The Heisenberg length scale can be identified, for
instance, by varying $J$ and correspondingly fixing $\gamma =
\gamma_c(J)$. Along this line vector observables computed from
$G_V({\bm x})$ would scale in terms of $W = (J - J_c) L^{1/\nu_H}$,
i.e., with the Heisenberg length-scale exponent.

\section{Conclusions}
\label{conclu}

In this paper we address the nature of the so-called O($N$)$^*$
  and Ising$^*$ universality classes, that differ from the standard 3D
  O($N$) vector and Ising universality classes because of the absence
  of critical vector correlations. These transitions occur for example
  in generalized lattice ${\mathbb Z}_2$-gauge models, i.e. the 3D
  ${\mathbb Z}_2$-gauge $N$-vector models, where all gauge-invariant
  quantities develop critical behaviors analogous to those of the
  standard $N$-vector or Ising (for $N=1$) model, without exposing
  critical order-parameter vector correlations. The fundamental spin
  field is not gauge invariant and therefore its correlation functions
  are trivial. This apparently precludes interpreting these
transitions as the result of the condensation of a vector order
parameter that breaks the O($N$) symmetry (in the gauge model the
global symmetry group is SO($N)$) down to O($N-1$), as in the standard
$N$ vector model.  O($N$)$^*$ transitions occur in 3D ${\mathbb
  Z}_2$-gauge $N$-vector models with Hamiltonian~(\ref{ham}), along
the line between the spin disordered and the spin ordered phase for
sufficiently large values of the inverse gauge coupling $K$, see
Fig.~\ref{phadiaN}.  Analogous Ising$^*$ transitions, see
Fig.~\ref{phadiaz2}, occur in the one-component model ($N=1$), which
is also known as the 3D ${\mathbb Z}_2$-gauge Higgs model.

We extend the characterization of the O($N$)$^*$ transitions, showing
that one can define a proper gauge-fixing procedure, which does not
change gauge-invariant correlations, but allows one to define vector
correlations that behave as in the $N$-vector model.  For this purpose
we propose a gauge-fixing procedure, which we name stochastic gauge
fixing, that is quite different from the usual gauge-fixing procedures
that are used in field theory. Here, gauge-dependent quantities are
averaged over gauge transformations weighted by an ancillary
gauge-dependent Hamiltonian. This leads to an extended quenched model,
which can be interpreted as a random-bond Ising model, in which the
bonds are distributed according to the Gibbs weight associated with
the original gauge-invariant Hamiltonian.  The random-bond Ising model
we consider differs from those typically studied in the random-system
literature. Indeed, one typically considers spatially uncorrelated
bond distributions, see, e.g.,
Refs.~\cite{Harris-74,EA-75,Nishimori-81,LH-88,PC-99,Betal-00,Nishimori-book,
  KKY-06,HPPV-07,HPV-08,CPV-11,Janus-13,LPP-16}.  Instead,
O($N)^*$/Ising$^*$ transitions separates two phases characterized by
the fact that the ${\mathbb Z}_2$-gauge variables show a topological
order.

It is important to remark that, if the ancillary Hamiltonian is chosen
such that the gauge-fixed vector correlations are critical, then their
critical behavior should necessarily coincide with that of vector
correlations in the standard O($N$) universality class (spin-spin
correlations for $N=1$), given that all gauge-invariant
observables---for instance, the spin-two operator $Q_{\bm x}^{ab}$
defined in Eq.~\eqref{qab} or the cumulants of the gauge-invariant
energy---behave as in the $N$-vector model. By using the stochastic
gauge fixing, we are thus able to identify the missing critical vector
fields, obtaining a full correspondence between O($N)^*$/Ising$^*$ and
standard O($N$)/Ising universality classes.

In this work, we define a simple ancillary system that makes
gauge-fixed vector correlations critical. Explicit numerical results
are obtained for $N=3$ and $N=1$.  We show that the stochastic gauge
fixing procedure allows us to observe the missing vector field which
orders at the transition, characterizing the breaking of the O($N)$
global symmetry.  Analogously, in the ${\mathbb Z}_2$-gauge model, we
identify the scalar field that breaks the global ${\mathbb Z}_2$
symmetry, which is absent in the gauge-invariant model (it is gauged)
but is restored by the gauge-fixing procedure.

Note that the use of the stochastic gauge fixing in the
  $\mathbb{Z}_2$-gauge Higgs model may also turn out to be convenient
  from the purely numerical point of view, since it simplifies (and
  likely makes more accurate) the analyses required to investigate the
  critical behavior of the model: In the gauge invariant model only
  cumulants of the energy can be studied, while in the gauge fixed
  model we have access to all the standard observables commonly used
  to study magnetic transitions.

The results presented in this paper show that the O($N$)$^*$
transitions in ${\mathbb Z}_2$-gauge $N$-vector models can still be
described by an effective O($N$)-symmetric LGW $\Phi^4$ theory,
without gauge fields.  However, the fundamental field in the effective
theory is not gauge invariant.  We remark that this behavior shows
some similarities and also significant differences with respect to the
one emerging at the Coulomb-Higgs transitions in the lattice AH models
with noncompact U(1) gauge variables, which are associated with a
stable charged fixed point of the AH field theory, see, e.g.,
Refs.~\cite{HLM-74,BPV-22}.  At variance with O($N$)$^*$ transitions,
at Coulomb-Higgs transitions gauge modes play a fundamental role and
therefore cannot be integrated out.  On the other hand, as in
O($N$)$^*$ transitions, the fundamental vector field is not
gauge-invariant and thus critical vector correlations are only
observed when the Lorenz gauge fixing is used, or, equivalently, if
one considers nonlocal gauge-invariant charged
operators~\cite{BN-87,BPV-23-chgf,BPV-24-ncAH}.

It is finally important to remark that the stochastic
  gauge-fixing method introduced in this work to study the O($N$)$^*$
  transitions of the $\mathbb{Z}_2$-gauge $N$-vector model can be
  easily generalized and extended to generic statistical models
  undergoing O($N$)$^*$ transitions in the presence of an emerging
  local discrete gauge symmetry. Actually, this approach can be
  straightforwardly extended to continuous gauge groups, such as the
  U(1) group. However, its utility within this extended context of
  continuous gauge groups must be checked, to understand whether it
  allows us to expose further features arising from gauge-dependent
  modes.  This may provide a novel way of studying observables whose
  form is very complicated (and in general non local) when written in
  a manifestly gauge invariant form, going beyond the techniques which
  use standard gauge fixing approaches~\cite{KK-85, KK-86,
    BPV-23-chgf, BPV-24-ncAH, BPV-23-gf}.

\acknowledgments

The authors acknowledge support from project PRIN 2022 ``Emerging
gauge theories: critical properties and quantum dynamics''
(20227JZKWP).  Numerical simulations have been performed on the CSN4
cluster of the Scientific Computing Center at INFN-PISA.

\end{document}